\documentclass[preprint,aps,amssymb]{revtex4}

\usepackage{graphicx}
\usepackage{epsfig}
\usepackage{longtable}
\usepackage{longtable}

\begin{document}

\title{Candidates for three-quasiparticle $K$-isomers in even-odd Fm  -  Cn nuclei}

\author{P.~Jachimowicz$^{1}$}
\author{M.~Kowal$^{2}$} \email{michal.kowal@ncbj.gov.pl}
\author{J.~Skalski$^{2}$}

\affiliation{$^1$ Institute of Physics, University of Zielona G\'{o}ra, Z. Szafrana 4a, 65-516 Zielona G\'{o}ra, Poland}
\affiliation{$^2$ National Centre for Nuclear Research, Pasteura 7, 02-093 Warsaw, Poland}

\date{\today}

\begin{abstract}

 Following our study of possible $K$-isomers in odd-even Md-Rg nuclei, here
 we continue with searching for three-quasiparticle 1$\nu$2$\pi$ and 3$\nu$
  isomer candidates in even-odd Fm - Cn nuclei.
  We use the same approach to calculate energies of different nuclear
configurations using a microscopic-macroscopic model with the Woods-Saxon
 potential. We used two versions of pairing:
 quasi-particle BCS method and particle number projection formalism.
 The optimal deformations for both ground states
 and high-$K$ configurations are determined through a four-dimensional energy
 minimization process.
 We point out the most promising candidates for high-$K$ isomers and compare them,
 where possible, with existing experimental data.


\end{abstract}


\maketitle

\section{INTRODUCTION}

Recent progress in experimental techniques will soon make it possible to conduct more systematic studies of excited states in superheavy (SH) nuclei, and enable them in systems with $Z > 105$.
Among various excitations, isomeric high-$K$ states present a special
interest, providing some information on low-lying single-particle (s.p.)
orbitals and mechanisms of decay hindrance.
The already established presence of high-$K$ isomers and
characteristics of the rotational bands in the No--Rf region
\cite{Herzberg2001,Ketelhut2009,Greenlees2012} are consistent with results of realistic microscopic-macroscopic (MM) and mean-field models, which predict well-deformed axially symmetric shapes
and numerous high-$\Omega$ orbitals close to the Fermi levels in these nuclei.
In the present work, following the previous one~\cite{JKS2023} devoted to
 three-quasiparticle (3qp) high-$K$ isomers in odd-even nuclei, we present
 predictions for 3qp isomers in
even-odd Fm--Cn nuclei using the micro-macro (MM) model with the Woods-Saxon
potential, extensively studied in the region of heavy and SH nuclei.

Since all nuclear models of superheavy nuclei (SHN) are merely extrapolations
from better-known regions of the nuclear chart, and experimental data on the
structure of $Z \ge 100$ systems is still scarce, an appreciation as to the
correctness of the former or the agreement or disagreement of both is rather
fragmentary at present. It is even more so, as often the only certain experimental result is the isomer half-life with some bounds on its
spin and energy. Then the spin value and configuration assignments are
given partially based on some theoretical result.

While it can be said that predictions of (the candidates for) isomeric
high-$K$ states reflect mostly the structure of the s.p. spectra, one has to remember that these spectra depend on equilibrium deformations, and those
change, even if a little, from isotope to isotope.
In this situation, only the more prominent features of the predicted high-$K$ configurations qualify for the model vs experiment test.

In the context of many-q.p. excitations, the important feature of the present model is the distinct subshell gap in the neutron spectrum at $N = 152$ around fermium and nobelium, which seem necessary for realistic predictions of
$K$-isomeric states. Two other subshell gaps predicted by the model: at
$N = 162$ for neutrons and at $Z = 108$ for protons, not quite confirmed
experimentally yet, strongly influence the predictions presented here.
The very similar Woods-Saxon model was used in~\cite{Stefan,ParSob}, which
can be consulted for predictions concerning 1-q.p. proton and neutron
excitations in this region of nuclei.

Up to now, high-$K$ isomers have been identified in a few even-odd nuclei.
Low-energy, one-quasiparticle states were found in $^{253}$Fm
(at $\approx 350$~keV)~\cite{Antalic2011}, $^{255}$No
(at 240--300~keV~\cite{Bronis2022} or 225~keV~\cite{Kessaci2024}) and
$^{257}$Rf (at $\approx 75$~keV)~\cite{Berr2010}, interpreted as the
$\nu 11/2^-[725]$ structure. 3qp isomers were found in $^{251,253,255}$No
(one~\cite{Hess2006,Lop2022}, one~\cite{Lop2007,Streicher2010,Antalic2011}
and two~\cite{Bronis2022,Kessaci2024}, respectively)
and in $^{253,255,257}$Rf (one~\cite{Lop2022,Khuy2021}, two
\cite{Mosat2020,Chakma2023} and one~\cite{Qian2009,Berr2010,Riss2013},
respectively), and one supposedly 5qp isomer in $^{255}$No
\cite{Kessaci2024}.
Spins and energies of two 3qp isomers in $^{255}$No were reported in
\cite{Kessaci2024}, spins and energies for $^{255}$Rf were given in
\cite{Mosat2020,Chakma2023} and energy of the 3qp isomer in $^{257}$Rf in
\cite{Riss2013}.
Properties of known isomers in neighbouring even-even Fm, No, and Rf nuclei
have an impact on the proposed assignments of the quasiparticle structure.
An example is provided by the 0.28~s, $8^-$ isomer in $^{254}$No
\cite{Tandel2006,Hessberger2010,Clark2010}, the
structure of which is related to that of isomers in $^{251,253}$No
and $^{254,255}$Rf~\cite{David2015}.
At the moment, the question of whether it is a two-proton or
two-neutron structure is unsettled~\cite{Lopez2}.
The recent collection of data on isomers in the heaviest nuclei is given
in~\cite{Ackermann2024,Ackermann2025}, see also
Ackermann~et~al.~\cite{Ackermann2015}, Asai~et~al.~\cite{Asai2015},
Dracoulis~et~al.~\cite{Dracoulis2016}, Walker~et~al.~\cite{Walker2020}, and A.~Lopez-Martens with K.~Hauschild~\cite{Lopez2}.

We search for $K$-isomers by calculating equilibrium energies of many
high-$K$3qp configurations, both $1\nu2\pi$ and $3\nu$, formed from
low-lying quasiparticle states. The set of studied 3qp states contains
 traditionally considered high-$K$ configurations called ``optimal''
 (i.e., obtained by the tilted Fermi surface method), but is considerably
 larger. For a theoretical overview based on the Nilsson-Strutinsky approach,
 see the work by Walker~et~al.~(2016)~\cite{Walker2016}.
As in the study on odd-even nuclei, we consider variants of pairing treatment:
 the quasi-particle method, and particle number projection that avoids
 deficiencies of the former, particularly for 3-neutron q.p. excitations.
In the present work, we studied nuclei in the following range of neutron numbers: $N = 141$--167 for Fm, No, Rf, Sg; $N = 143$--169 for Hs; $N = 145$--171 for Ds; and $N = 147$--173 for Cn.

The calculations and selection of candidates for high-$K$ isomers are described in Sect.~II, and results are presented and discussed in Sect.~III. The conclusions are given in Sect .~IV.

\section{THE METHOD}

 We use the MM method with the deformed Woods-Saxon (WS) potential
 \cite{Cwiok1987} and the macroscopic Yukawa-plus-exponential energy model
 \cite{Krappe1979} with parameters specified in \cite{Muntian2001}.
 Parameters of the model are kept the same as in all recent applications
 to heavy and superheavy nuclei, which concern masses and deformations
 \cite{Kowal2010}, $Q_\alpha$ - energies \cite{Jachimowicz2014}, the first
 and second fission barriers in actinides \cite{Jachimowicz2012-20} and SH
 nuclei \cite{Jachimowicz2017-2}, \cite{Jachimowicz2021}.
In order to obtain ground-states and configuration-constrained minima, we use a four-dimensional space of deformations $\beta_{\lambda 0}$ defining the nuclear surface:
\begin{equation}
\label{radius}
R(\theta)= c (\beta) R_0 \left[1+\sum_{\lambda=2,4,6,8}\beta_{\lambda 0} {\rm Y}_{\lambda 0}(\theta)\right],
\end{equation}
where ${\rm Y}_{\lambda 0}(\theta)$ are spherical harmonics, $c(\beta)$ is the
 volume-fixing factor depending on deformation, and $R_0$ is the
 radius of a spherical nucleus.
 Deformation parameters beyond $\beta_{20}$ and $\beta_{40}$ turned out
 necessary in studies of g.s. and high-K configuration equilibria in very heavy
 nuclei \cite{Patyk19911,Patyk19912},\cite{Liu2011}.
 On the other hand, calculations within our MM model \cite{Jachimowicz2021}
 showed that the reflection-asymmetric axial equilibrium deformations of
 the considered nuclei are either zero or negligible, so their omission
 is substantiated. Therefore, the intrinsic parity of the considered states is well-defined.

 Four-dimensional energy minimization over $\beta_{20}, \beta_{40}, \beta_{60},
 \beta_{80}$ (\ref{radius}) was performed using the gradient method. To avoid
 secondary or very deformed minima, the minimization is repeated at least 10 times for each configuration with different starting values of deformations.

 In the previous work, dealing with candidates for 3qp isomers in heavy
 odd-even nuclei, we have found that blocking two, and especially three
 quasi-particles in the BCS procedure (blocking approximation) produce a
 too small or vanishing pairing gap and too low 3qp excitation energies.
 Therefore, in the present study, we use exclusively the quasi-particle
 method and the particle-number-projection method (PNP).



 In the quasiparticle method, the microscopic part of the energy for a 1$\nu$2$\pi$ configuration is taken as the sum of BCS quasi-particle energies of singly occupied levels:
\begin{eqnarray}
\label{quasi}
E_{q.p.}^{*} = \sqrt{(\epsilon_{\nu}-\lambda_{\nu})^2+\Delta_{\nu}^2}
   + \sqrt{(\epsilon_{\pi_1}-\lambda_{\pi})^2+\Delta_{\pi}^2}   \nonumber
   + \sqrt{(\epsilon_{\pi_2}-\lambda_{\pi})^2+\Delta_{\pi}^2}     \nonumber  \\
\end{eqnarray}
and the core energy term consisting of the shell and pairing corrections
 calculated without blocking.
 The total, i.e., including the macroscopic part, energy, is then
 minimized over deformations.
 For neutrons, the core term, as well as the pairing gap $\Delta_{\nu}$ and the Fermi energy $\lambda_{\nu}$, are calculated for the odd number of particles, but with the double occupation of all levels.
 This prescription was used before in \cite{ParSob}.
 For 3$\nu$ q.p. excitations, the microscopic energy was the sum of three
 neutron-quasiparticle energies and the shell and pairing corrections
 calculated without blocking.
 As for the 1$\nu$2$\pi$ excitations, the odd
 particle number and double occupation of levels were used in the BCS
 procedure for neutrons. In the quasiparicle method, we use the same pairing strengths as in our mass model ($I=(N-Z)/A$):
  $ G_p = (g_{0p} + g_{1p}I)/A$ for protons and $G_n = (g_{0n} + g_{1n}I)/A$ for neutrons, with $g_{0p} = 13.40$ MeV , $g_{1p} = 44.89$ MeV,  $g_{0n} = 17.67$ MeV , $g_{1n}=-13.11$ MeV.

 One should mention that the quasiparticle method, especially for
 $3\nu$-q.p. excitations, has several flaws. First, it underestimates
 excitation energies at particle numbers for which the BCS energy gap vanishes or is minimal. This happens in the vicinity of substantial gaps in the s.p. spectrum, for odd $N=N_{gap}\pm 1$, with $N_{gap}$ the number of neutrons on levels below the gap. Moreover, if two additional levels are blocked above the gap for $N=N_{gap}+1$, or below the gap for $N=N_{gap}-1$, their excitation energies are underestimated further.
  In the former case, the Fermi energy of the core (accounting for half of
  the BCS occupancy of the level above the gap) lies just below the upper
  edge of the gap. Blocking two additional levels above the gap requires that
  the Fermi level of the $N_{gap}-2$ {\it paired} particles be lowered to
  ensure the correct (expectation value of the) particle number.
  In the latter case, the Fermi level of the core lies just above the lower
  edge of the gap while the Fermi energy of the $N_{gap}-4$ paired particles
  should be elevated as two of them must be promoted above the gap.
  The releted increase in energy of the 3qp states is not reflected in the sum
  of quasi-particle energies (\ref{quasi}) as the Fermi energy of the core is
  close to the s.p. energies of two additional blocked levels.

 In order to correct for the related error, one could adjust the Fermi level
 of the core for each set of blocked levels to obtain the correct
 number of paired particles $<{\hat N}_{pair}>=N-3$ on unblocked levels
 while keeping $\Delta_{\nu}$ of the core unchanged (ignoring the second BCS equation).
 We have checked that this gives a correction in the right
 direction. On the other hand, such a procedure does not seem worthwhile
  in comparison to the PNP solution, so we did not pursue it further.
 We have also made a second 3$\nu$ q.p. calculation with the core taken as
 the system of $N-1$ (even number of) particles. While both sets of results
 have some places in which the abovementioned errors occur, the second one
 produces more physical results for isotones $N=N_{gap}+1$.

 We use the Particle Number Projection (PNP) in order to correct for known
 deficiencies of the Bardeen-Cooper-Schrieffer (BCS) method in the weak pairing
 regime. The procedure is to select the component of the BCS-like state
 $\Psi(\lambda,\Delta)$ (with $\Delta$ and $\lambda$ unrestricted by the BCS
 equations) with exactly $n$ pairs of particles, and calculate on it the
 expectation value $E_n$ of the pairing Hamiltonian.
 This should be done independently for the neutron and proton subsystems
 ($2n=Z$ or $Z-2$ for g.s. or $2\pi$ q.p. states, and $N-1$ or $N-3$ for
  $1\nu$ or $3\nu$ q.p. states).

  In the majority of calculations, we used the well-known method of Fomenko
  \cite{Fomenko1970}. We also implemented a second method for the purpose of
   tests, which consists in expressing $E_n$ in terms of norms of
  particle-projected BCS-like states:
\begin{equation}
\label{Erzut_alt}
 E_n=\frac{\sum_{\mu>0}(2\epsilon_{\mu}-G)v^2_{\mu}P^{\mu}_{n-1} -
    G\sum_{\mu>0\ne\nu>0}u_{\mu}v_{\mu}u_{\nu}v_{\nu}P^{\mu \nu}_{n-1}}
    {P_n}  ,
\end{equation}
 where $P_n$, $P^{\mu}_{n-1}$ and $P^{\mu \nu}_{n-1}$ are the squared norms of:
 the $n$-pairs component of $\Psi(\lambda,\Delta)$, the $n-1$ pairs component
 of $\Psi(\lambda,\Delta)$ with the level $\mu$ excluded,
  and the $n-1$ pairs component of $\Psi(\lambda,\Delta)$ with excluded levels
 $\mu$ and $\nu$, respectively; $\epsilon_{\mu}$ are s.p. energies, $G$ -
  the pairing strength, $u_{\mu}$ and $v_{\mu}$ - BCS-like amplitudes.
 Since $P_n$ is the coefficient by $\zeta^{n}$ of the polynomial
 $\prod_{\nu>0} (u^2_{\nu}+\zeta v^2_{\nu})$, one can find it
 by representing $\zeta$ as the $(n+1)$-dimensional Jordan block with zeros on
 the main diagonal and one on the first superdiagonal and calculating
  the element $(n+1, n+1)$ of the matrix product
 $\prod_{\nu>0} (u^2_{\nu}\ \mathbb{I} +\zeta v^2_{\nu})$.
 Quantities $P^{\mu}_{n-1}$ can be obtained in the same way, and from them
 $P^{\mu \nu}_{n-1}$ - see Appendix of \cite{JKS2023}.
 Both computational schemes were checked to give
 identical results for $E_n$.

 To simplify the search for the optimal projected BCS-like state into a
 single-parameter minimization, parameter $\lambda$ of $\Psi(\lambda,\Delta)$
 is fixed by requiring that the expectation value of the particle number
 operator, $\langle \Psi(\lambda,\Delta)\mid{\hat N}\mid
 \Psi(\lambda,\Delta)\rangle$, be equal the desired number of
 particles. The resulting projected energy is then a function of
 $\Delta$ alone and exhibits a unique minimum. This minimal value is adopted as the energy after projection.
 For each configuration, the minimization of $E_n$ over $\Delta$ is
 performed at each step of the minimization over deformation.

 In the previous work \cite{JKS2023}, we have chosen pairing strengths
 $G_n(N,Z)$ for neutrons and $G_p(N,Z)$ for protons in the
 PNP procedure as: $G_n(N,Z)=1.1\ G_n^{mod}(N,Z)$, $G_p(N,Z)=1.1\
 G_p^{mod}(N,Z)$, with $G_n^{mod}(N,Z)$, $G_p^{mod}(N,Z)$ the pairing strengths from our MM model. With this choice, we obtained the energy of the 2-q.p. neutron excitations involving
 two nearly degenerate s.p. levels at the Fermi energy: the last occupied and
 the first empty one, close to $2\Delta_n({\rm BCS})$ of our MM model, while
 the proton strengths ratio $G_p/G_p^{mod}$ was chosen equal to that for
 neutrons.
 In the present calculations, we first used the same parameters.
 After noticing that the smallest excitation energies of $3\nu$ q.p. states
 of $\approx 0.65$ MeV are clearly too small, we made a second calculation
 with $G_n(N,Z)=1.15\ G_n^{mod}(N,Z)$.
 The pairing strength for protons was kept the same, as it has a negligible effect on $3\nu$ q.p. state energies. As a result, the smallest energy of the
 $3\nu$ q.p. states increased to $\approx 1$ MeV, which looks more realistic.
 One can notice that the neutron pairing strength $G_n$ practically
 does not influence energies of the $1\nu2\pi$ q.p. states.

 Equilibrium deformations were found by the quasiparticle method for 6012
 $1\nu2\pi$ configurations with $K_{\pi}=\Omega_{\pi 1}+\Omega_{\pi 2}\ge 5$
 and  5824 $3\nu$ q.p. configurations with
 $K_{\nu}=\Omega_{\nu 1}+\Omega_{\nu 2}+\Omega_{\nu 3} \ge 11/2$
 built from s.p. states not too distant from the Fermi surfaces in the studied nuclei. Using these results as a guidance, the PNP calculations were confined to some selected low-lying 3qp states.
Configurations at the lowest excitation energies $E^*_{3q.p.}(K)$ are
considered as likely candidates for $K$-isomers. The angular momentum $I$ of
 the 3q.p. configuration is understood as a sum of s.p. angular momentum
  projections $\Omega_i$. We did not consider
 shifts due to the spin interaction (counterpart of Gallagher shifts for
2q.p.), which for 3q.p. configurations are not well understood
\cite{Pyatov1964,Jain1992}.

 As mentioned in our study of odd-even systems \cite{JKS2023}, an additional indicator of possible isomerism of the high-$K$3-q.p. configuration may be it's excitation above the state of the same spin of the collective  rotational
 band built on its 1$\nu$ q.p. component (more loosely, its excitation above
 the yrast line). The rotational energy of the 1$\nu$ collective band with
 $K_{\nu}=\Omega_{\nu}$ and the band-head at the excitation energy
 $E_{1q.p.}^*(\Omega_{\nu})$ (wich would be zero for the g.s.) can be estimated
 as: $E^{rot}(I=K,\Omega_{\nu})=\frac{1}{2}[K(K+1)-\Omega_{\nu}^2]/{\cal J}$,
 with ${\cal J}$ - average moment of inertia of the 1-neutron q.p. rotational
 band. The smaller the value $[E^*_{3q.p.}(K)
 -E_{1q.p.}^*(\Omega_{\nu})-E^{rot}(I=K,\Omega_{\nu})]$, the less probable
 is the high-$K$ 3 q.p. state deexcitation to the one-neutron q.p.
  rotational band. Thus, for 1$\nu2\pi$ states, a large extra-aligned
  $K_{2\pi}=K_{\pi 1}+K_{\pi 2}$ of the two-proton pair is favourable for
  the isomer existence. For a $3\nu$ configuration, decays are possible to
 three rotational bands built on each $1\nu$ component. The most
 probable of them would limit the isomer existence. In particular,
 if $K_{\nu 1}$ and $K_{\nu 2}$ are much smaller than $K_{\nu 3}$, and
 three excitation energies of $1\nu$ band-heads are similar,
 the stringest condition for $3\nu$ isomer would be a decay to the
 rotational band built on $K_{\nu 3}$ - its rotational energy $E^{rot}$
 is the smallest as it corresponds to the smallest collective angular momentum
  $K_{\nu 1}+K_{\nu 2}$ (equal or similar moments of inertia are tacitly
  assumed for all $1\nu$ rotational bands).

 As in \cite{JKS2023} we used the calculated cranking moments of inertia
 of even-even nuclei from \cite{momJ} to estimate rotational energies in
 studied isotopes. For odd-$N$ nuclei, we took the average from the
 calculated moments of inertia in neighboring even-even nuclei and
 increased it by a factor 1.4.

\section{RESULTS AND DISCUSSION}



\begin{figure}[h]
\centering
\includegraphics[width=0.55\textwidth]{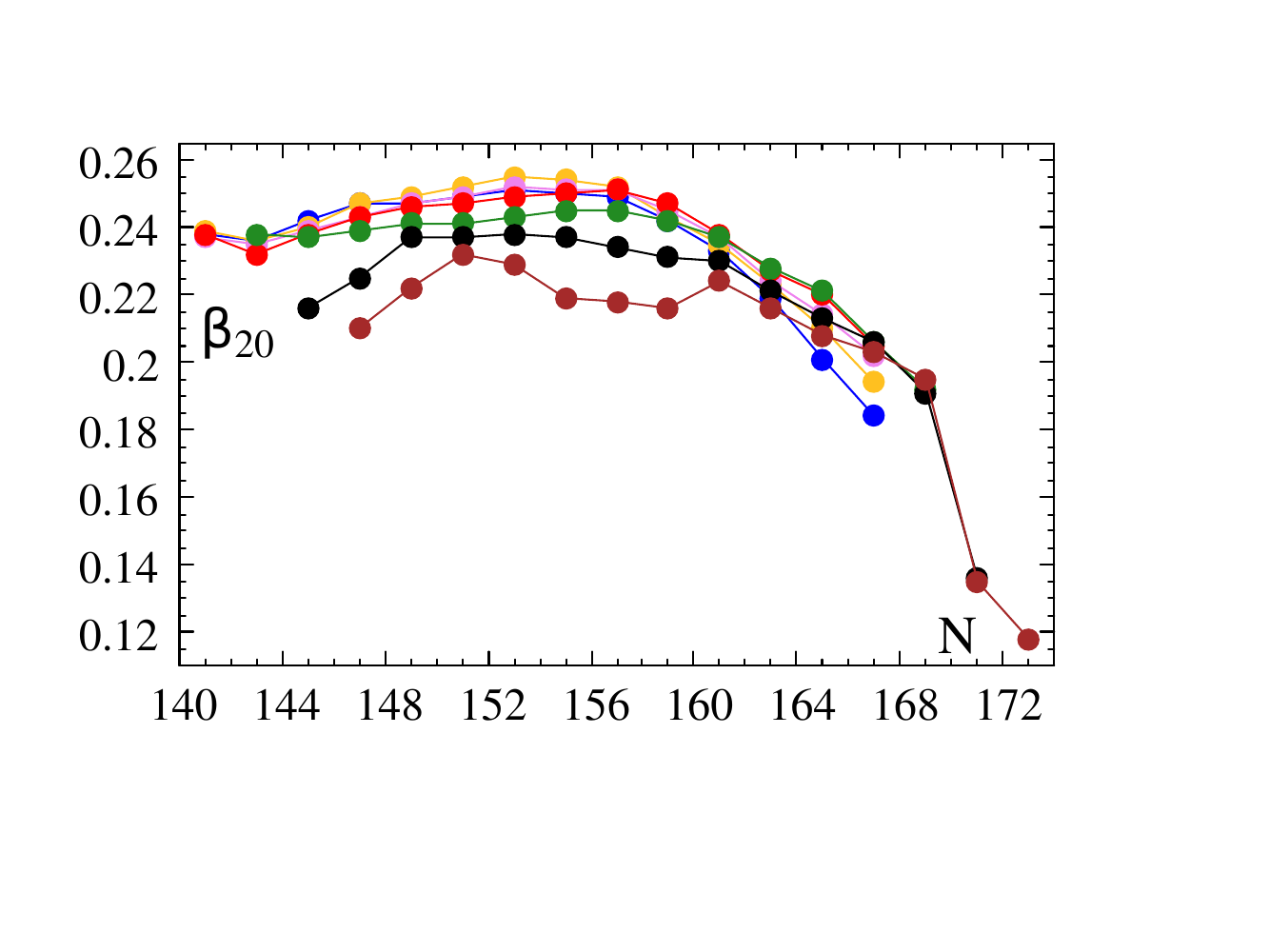}
\hspace{-15mm}
\includegraphics[width=0.55\textwidth]{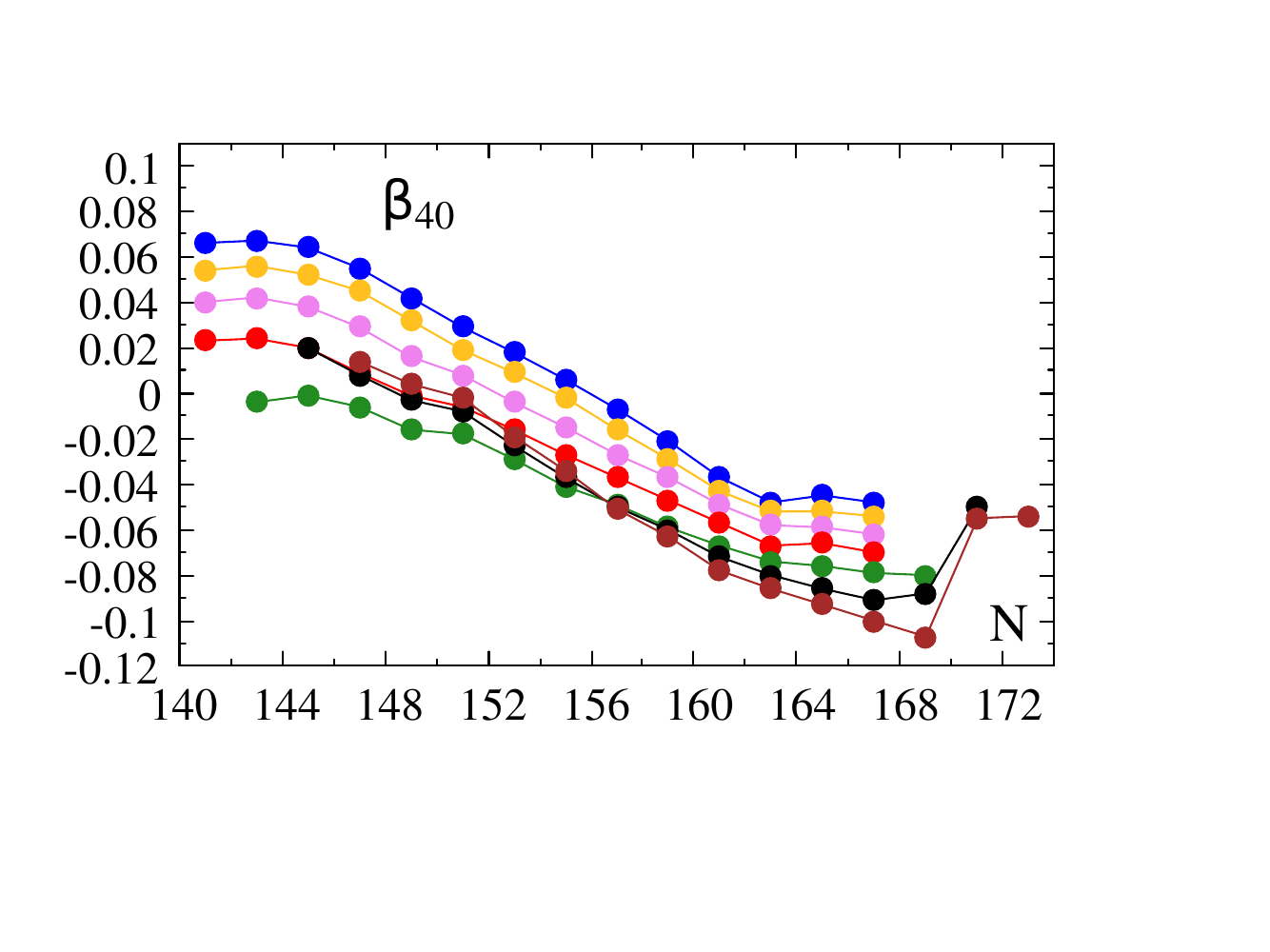}\\
\vspace{-20mm}
\includegraphics[width=0.55\textwidth]{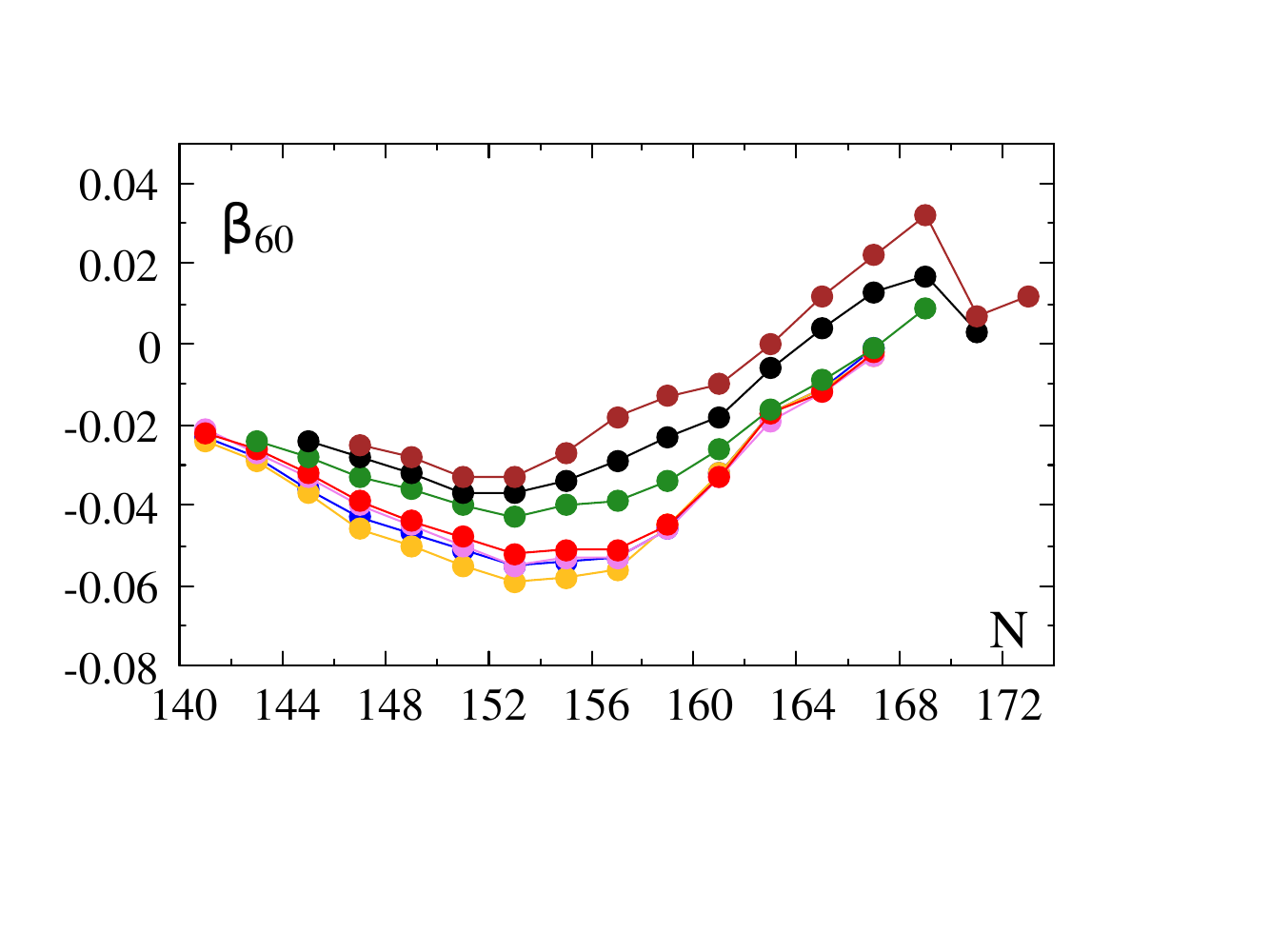}
\hspace{-15mm}
\includegraphics[width=0.55\textwidth]{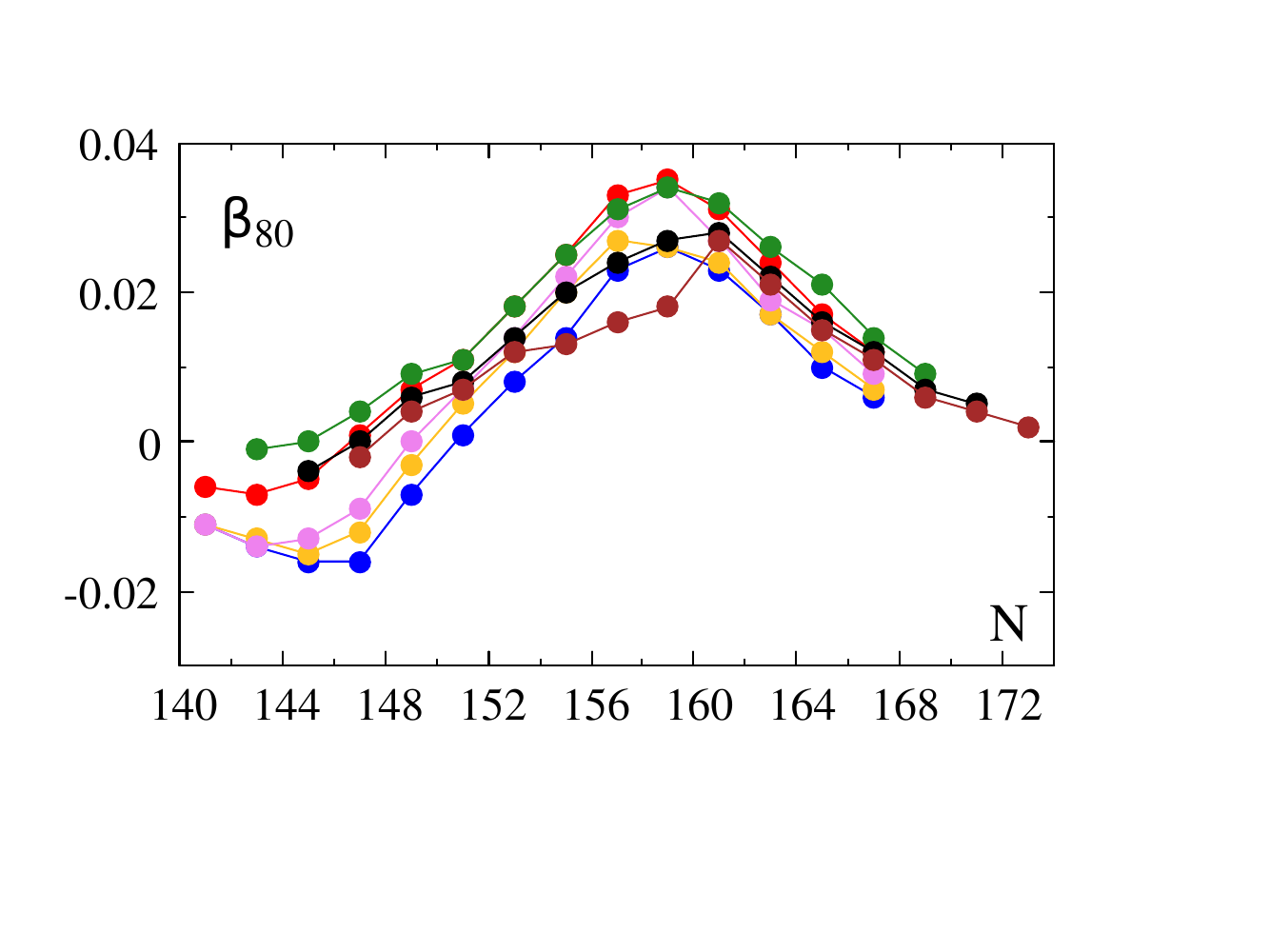}
\vspace{-15mm} 
\caption{Equilibrium g.s. deformations obtained in the MM
 model with the quasiparticle method for even-odd isotopes of: Fm - blue,
 No - yellow, Rf - violet, Sg - red, Hs - green, Ds - black, Cn - brown;
 Note different scales on the ordinate.
 }   
\label{eqdef}
\end{figure}

 At present, experimental information on even-odd Fm-Cn nuclei exists for
 neutron numbers between $N=$141-159 for Fm, $N=$147-157 for No,
 $N=$149-163 for Rf, $N=$153-163 for Sg, $N=$155-169 for Hs, $N=$157-171 for
 Ds and $N=$165-173 for Cn.

 Equilibrium deformations obtained for studied nuclei in our MM model with
 the q.p. method are shown in Fig. \ref{eqdef} (very similar deformations
 obtained with the blocking method are reported in \cite{Jachimowicz2021}).
 Deformation $\beta_{20}$ in Fm-Hs is mostly in the range 0.23-0.25, with a
 wide maximum around $N=$153-157, slightly decreasing towards 0.19 - 0.20
 for larger $N$; in Ds and Cn it is smaller, with a sharp decrease
 for $N=171$. Deformations $\beta_{40}$ mostly decrease with $N$ from a
 wide maximum at N=143 towards a value by $\approx 0.1$ smaller at $N=167$;
 they show a spread in $Z$, with larger values occuring mostly for
 smaller $Z$.
 Deformation $\beta_{60}$ shows a wide minimum around $N\approx 155$.
 It changes within a range of $\approx 0.06$ for each element, and
 at fixed $N$ it mostly increases with $Z$.
 Deformation $\beta_{80}$ is generally small, showing roughly a sinusoidal
 pattern as a function of $N$, with the values $\approx 0.02 - 0.03$ at the
 maximum around $N=158,160$.
 Equilibrium deformations for the majority of 3q.p. configurations are mostly
 close to those of the ground states, but the differences
  in $\beta_{20}$ can reach sometimes $0.02-0.03$.
 In the PNP method, equilibrium deformations are in most cases very similar
 to those in the q.p. method, with a tendency towards slightly smaller
 values.

\begin{figure}[!tbp]
\centerline{\includegraphics[scale=0.7]{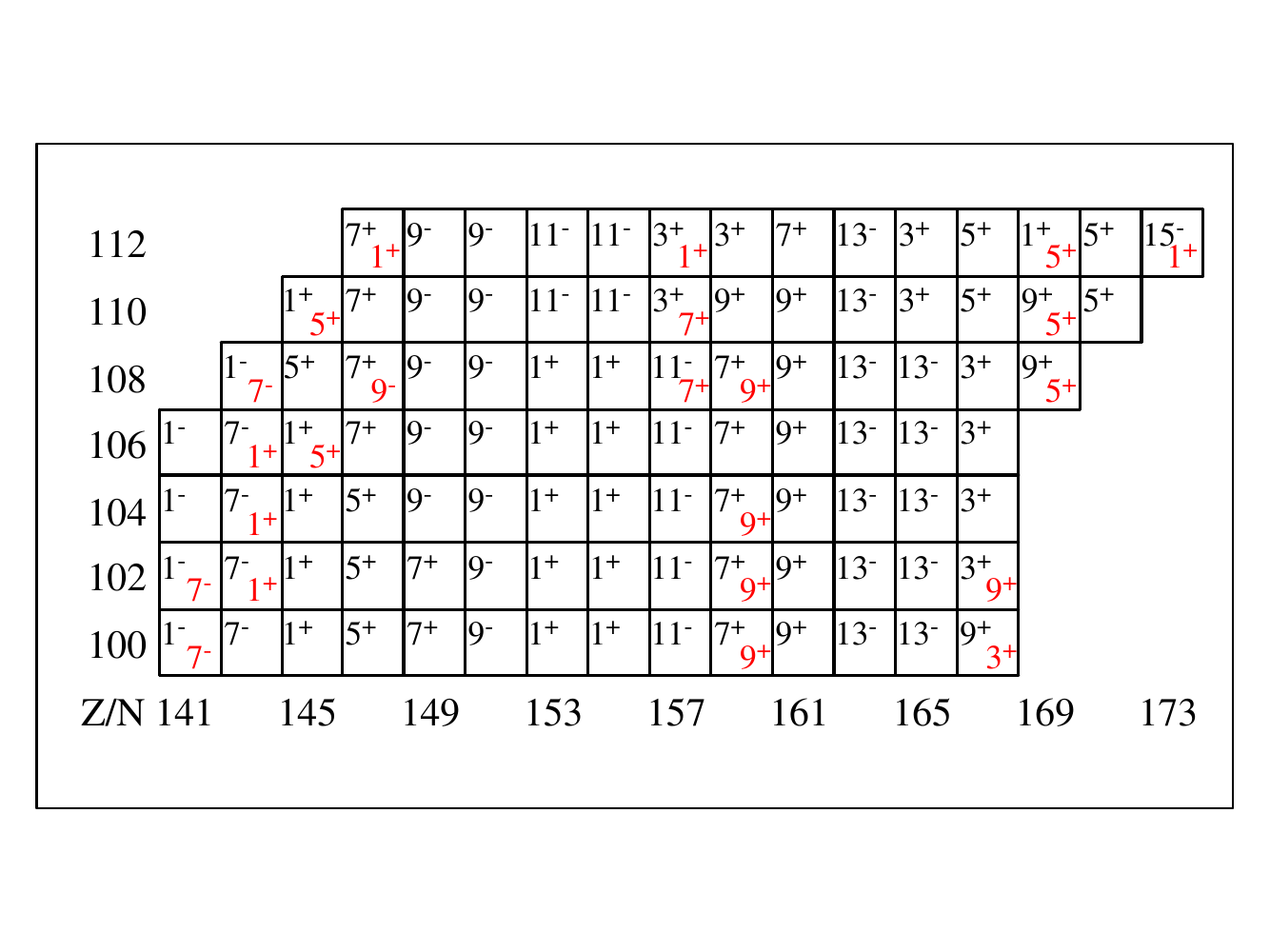}}
\caption{{\protect Ground-state spins and parities (given as
 $2\Omega^{\pi}$) of even-odd isotopes from our MM model
 calculated with the qusiparticle method (upper left, in black) and
  with the PNP (given only if different - lower right, in red).
 }}
\label{szach1nq}
\end{figure}

 Ground-state spins and parities of the considered odd-$N$ nuclei are
 usually (but not always) determined by the s.p. spectrum - properties
 of the singly occupied neutron orbital at the Fermi level.
 Results obtained in the quasiparticle scheme are shown in Fig.
 \ref{szach1nq}.
 The g.s. spins and parities from the PNP calculation (indicated in
 Fig. \ref{szach1nq} in red only if different from q.p. result) differ from
 the above only when the neutron Fermi level falls in the vicinity of nearly
degenerate neutron orbitals.

\begin{figure}[!tbp]
\centerline{\includegraphics[scale=0.6]{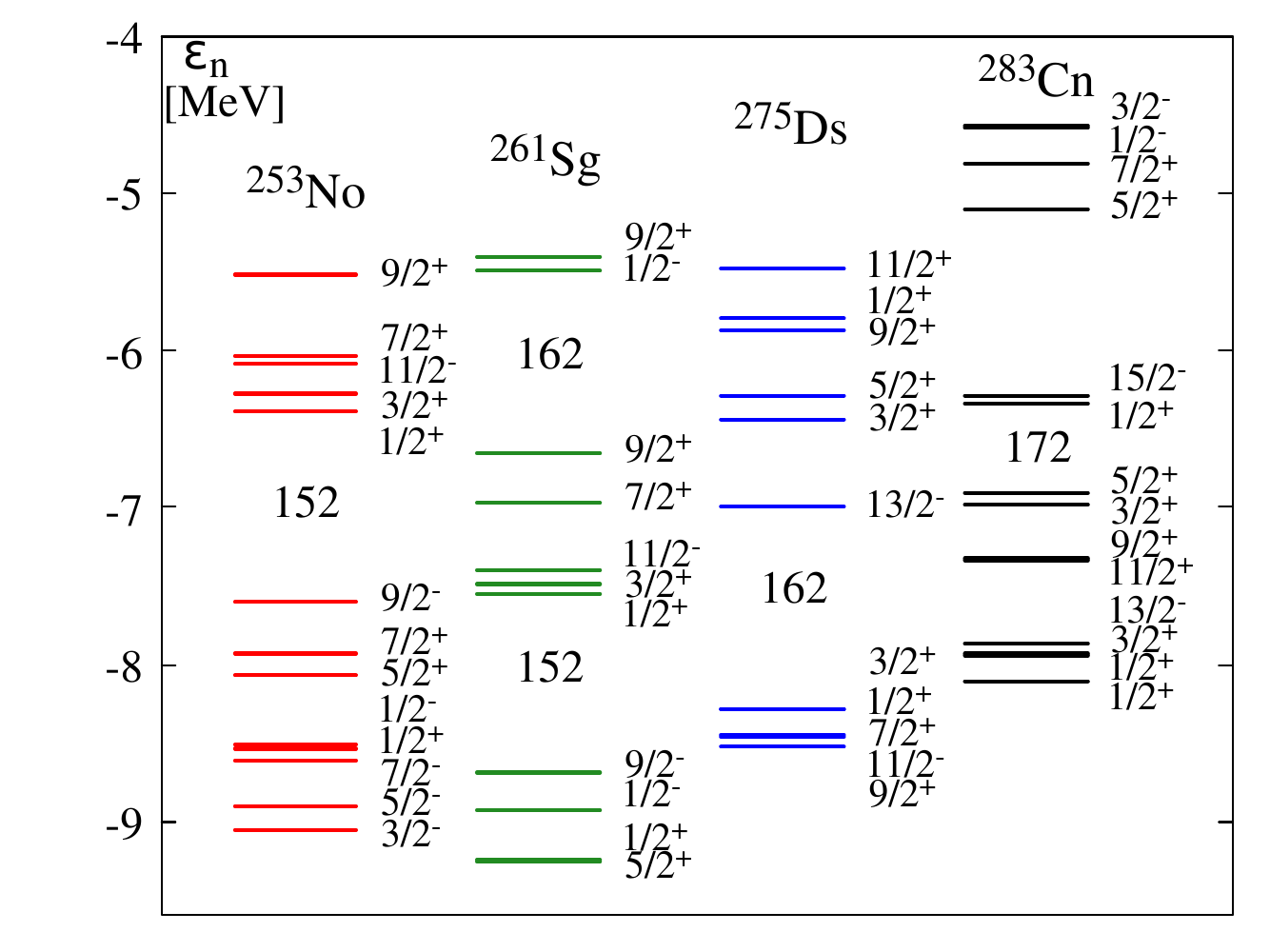}}
\caption{{\protect Single-particle neutron levels
 in the Woods-Saxon potential of our MM model for four nuclei at their
  g.s. equilibrium deformations calculated in \cite{Jachimowicz2021}.
 }}
\label{nlev2}
\end{figure}

 Characteristic of the s.p. neutron spectrum in our Woods-Saxon model
 are two gaps at $N=$152 and 162. Below the lower gap one has (from $N=141$ up)
 the following states: $1/2^-_{\sf 13}[501]$, $7/2^{-}_{\sf 5}[743]$,
 $1/2^+_{\sf 13}[631]$, $5/2^+_{\sf 7}[622]$, $7/2^{+}_{\sf 4}[624]$,
  and $9/2^{-}_{\sf 3}[734]$; between the gaps there are five orbitals:
 $1/2^+_{\sf 14}[620]$, $3/2^+_{\sf 10}[622]$, $11/2^{-}_{\sf 2}[725]$,
  $7/2^{+}_{\sf 5}[613]$ and $9/2^{+}_{\sf 3}[615]$; above the $N=$162 gap
  there are following states: $13/2^{-}_{\sf 1} [716]$,
  $9/2^{+}_{\sf 4}[604]$, $3/2^+_{\sf 11}[611]$ and $5/2^+_{\sf 8}[613]$.
  For neutron numbers $N=169,171,173$ in Hs, Ds, and Cn, the equilibrium
  deformations $\beta_{20}$ are significantly smaller and the following
  additional states are close to the Fermi level: $11/2^+_{\sf 2}[606]$,
 $1/2^+_{\sf 15}[611]$, $15/2^-_{\sf 1}[707]$
 (Single-particle (s.p.) states are denoted by $\Omega^{\pi}_{\sf n})$, where $\Omega$ is the projection of the s.p. total angular momentum onto the
 symmetry axis, $\pi$ is its parity, and the subscript `${\sf n}$' is the
 ordinal number of the state (with a given $\Omega^{\pi}$) counted from the
 lowest energy within that $\Omega^{\pi}$ block. Additionally, states may be
 identified by their asymptotic Nilsson quantum numbers $[N n_z \Lambda]$
 (where $N$ is the principal quantum number, $n_z$ the oscillator quanta along
the symmetry axis, and $\Lambda$ the projection of orbital angular momentum).
 In Fig. \ref{nlev2}, the calculated s.p. neutron levels at equilibrium
 deformations are shown for four nuclei: $^{253}$No, $^{261}$Sg, $^{275}$Ds
 and $^{283}$Cn.

\begin{figure}[!tbp]
\centerline{\includegraphics[scale=0.6]{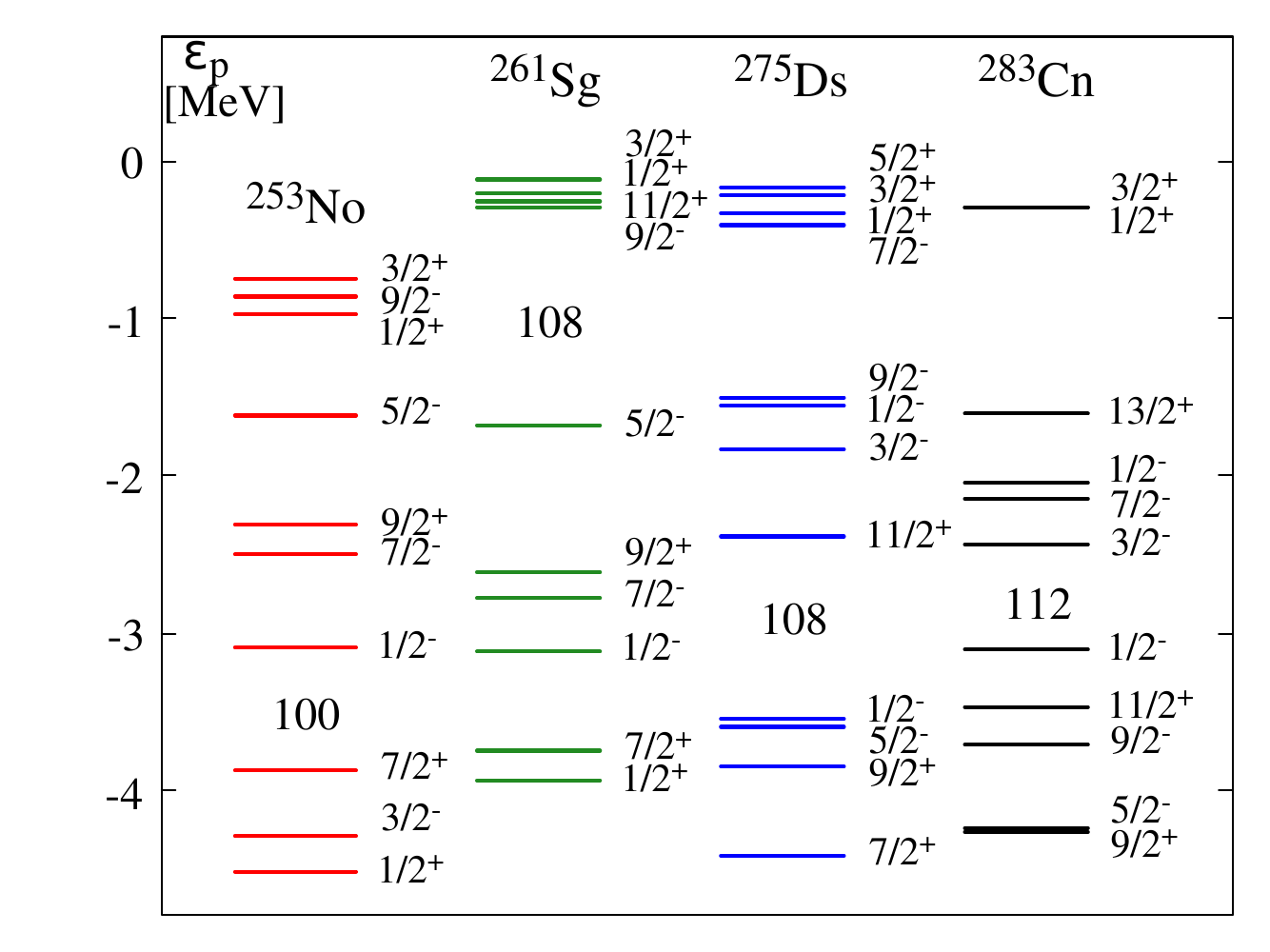}}
\caption{{\protect Single-particle proton levels
 in the Woods-Saxon potential of our MM model for four nuclei at their
 g.s. equilibrium deformations calculated in \cite{Jachimowicz2021}.
 }}
\label{plev2}
\end{figure}

 Experimentally established ground state spins and parities are:
 in odd-$N$ isotopes $^{249-257}$Fm: $7/2^+$, $9/2^-$, $1/2^+$, $7/2^+$
 and $9/2^+$, and in four nobelium isotopes $^{249,253,255,259}$No:
 $5/2^+$, $1/2^+$, $9/2^-$ and $9/2^+$ \cite{Kondev2021}.
 The g.s. configurations calculated within the PNP method are
 different in $^{255}$Fm and $^{257}$Fm, in which the experimental
 configurations come out excited, respectively, by 140 and 160 keV,
 and in $^{259}$No, in which the experimental configuration is obtained
 at the excitation of 170 keV. These comparisons suggest a slightly different
 order or spacing of neutron levels, in particular, a higher placement of the
 $\nu 11/2^-_{\sf 2}$ state and a lower one of the $\nu 7/2^+_{\sf 5}$ and
 $9/2^+_{\sf 3}$ states.
  The left-hand part of Fig. \ref{nlev2} shows the relative positions of the neutron levels above the $N=152$ gap, relevant for the above comparison.

\subsection{Excitation energies of $1\nu2\pi$ 3-q.p. \mbox{high-$K$}
 candidates for isomeric states}

\begin{figure}[!tbp]
 \includegraphics[width=0.8\linewidth]{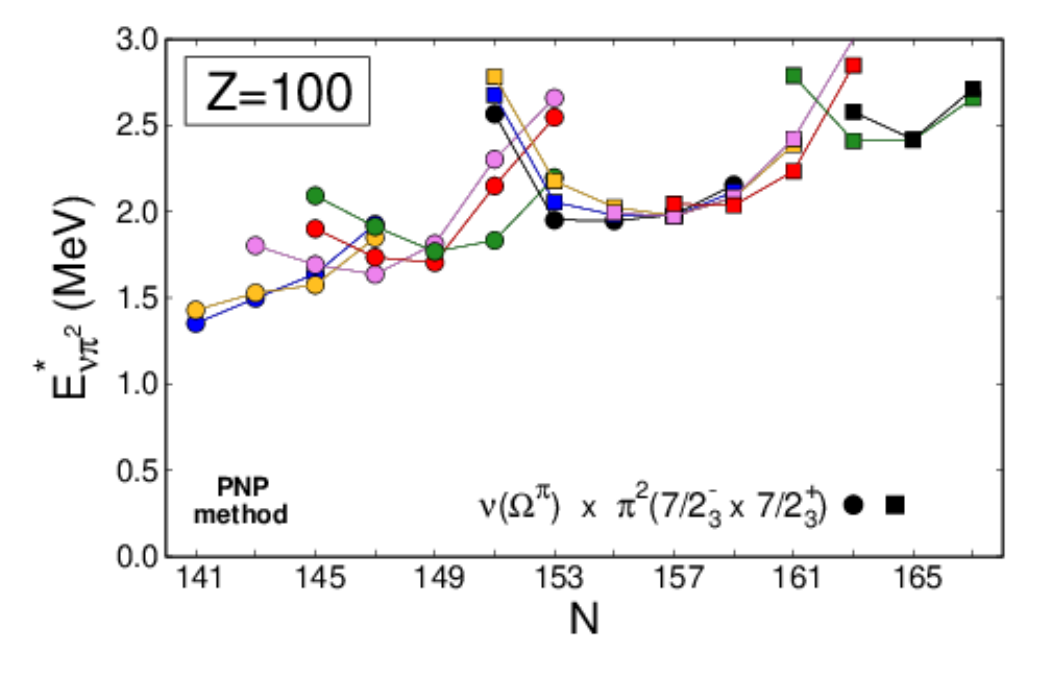}
\includegraphics[width=0.8\linewidth]{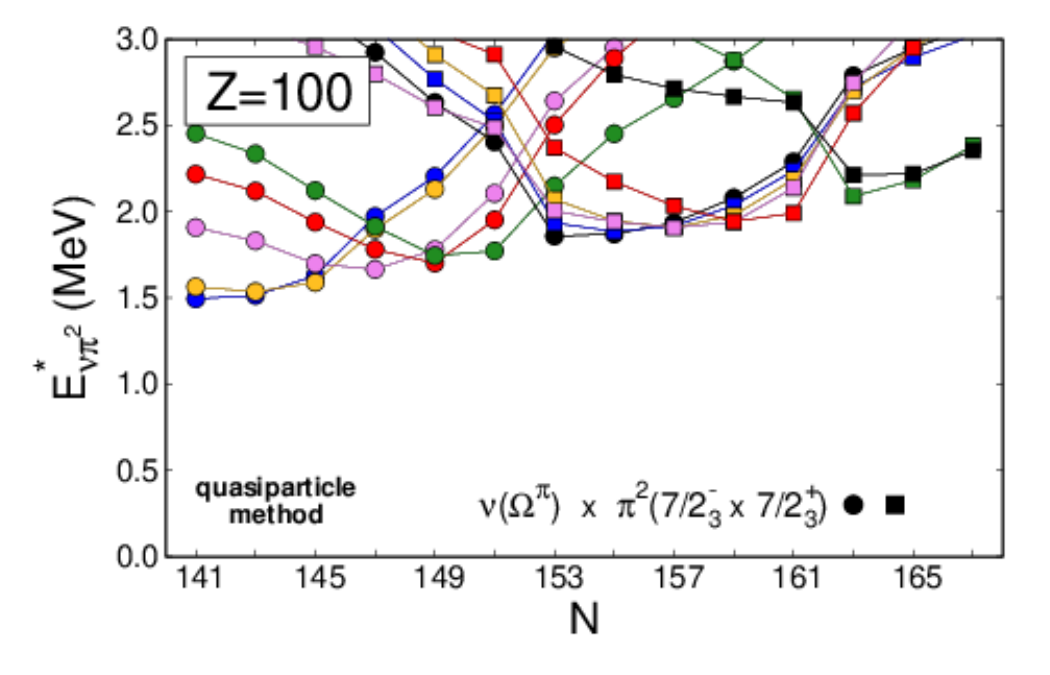}
\caption{Energies of low-lying $1\nu2\pi$ large-$K$ excitations in Fm isotopes
 vs $N$. The upper panel shows results of the PNP formalism, the lower one -
 of the quasiparticle method. Configurations composed of the favoured
 two-proton pair $\pi^2\{7/2^-_{\sf 3}[514]\otimes 7/2^+_{\sf 3}[633]\}$
 and the lowest neutron states $\Omega_{\nu}^{\pi}$ following the rising
 neutron Fermi level is distinguished by the neutron component:
 $7/2^-_{\sf 5}[743]$ - blue circle,  $1/2^+_{\sf 13}[631]$ - yellow circle,
 $5/2^+_{\sf 7}[622]$ - violet circle,  $7/2^+_{\sf 4}[624]$ - red circle,
 $9/2^-_{\sf 3}[734]$ - green circle,  $1/2^+_{\sf 14}[620]$ - black circle,
 $3/2^+_{\sf 10}[622]$ - blue square,  $11/2^-_{\sf 2}[725]$ - yellow square,
 $7/2^+_{\sf 5}[613]$ - violet square,  $9/2^+_{\sf 3}[615]$ - red square,
 $13/2^-_{\sf 1}[716]$ - green square,  $9/2^+_{\sf 4}[604]$ - black square.  }
\label{1n2p_100}
\end{figure}

\begin{figure}[!tbp]
\includegraphics[width=0.8\linewidth]{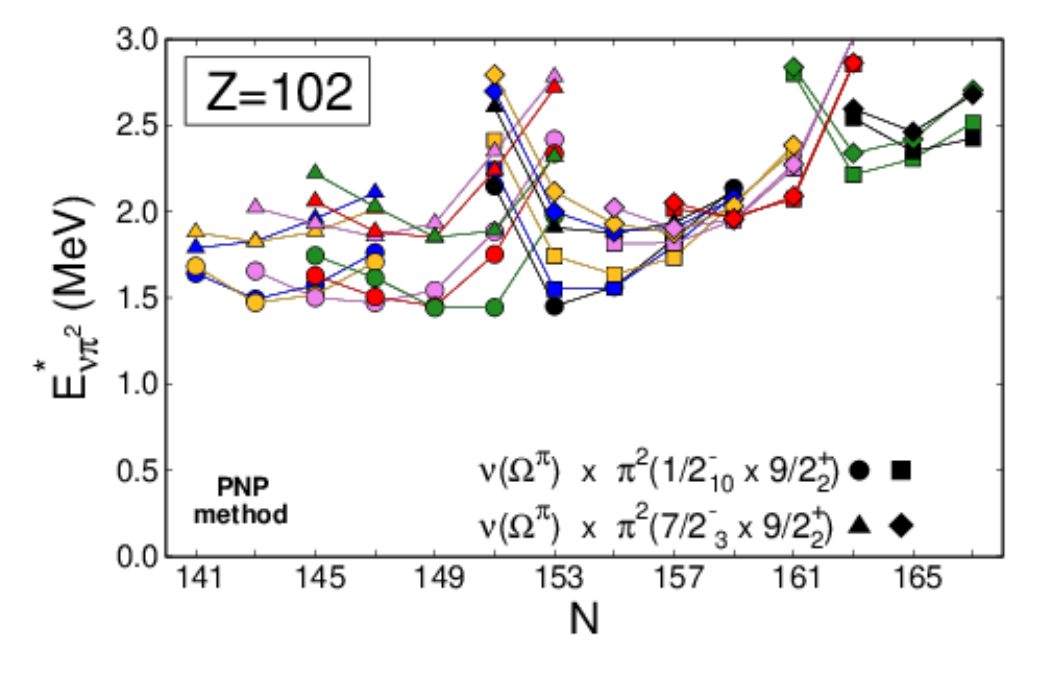}
\includegraphics[width=0.8\linewidth]{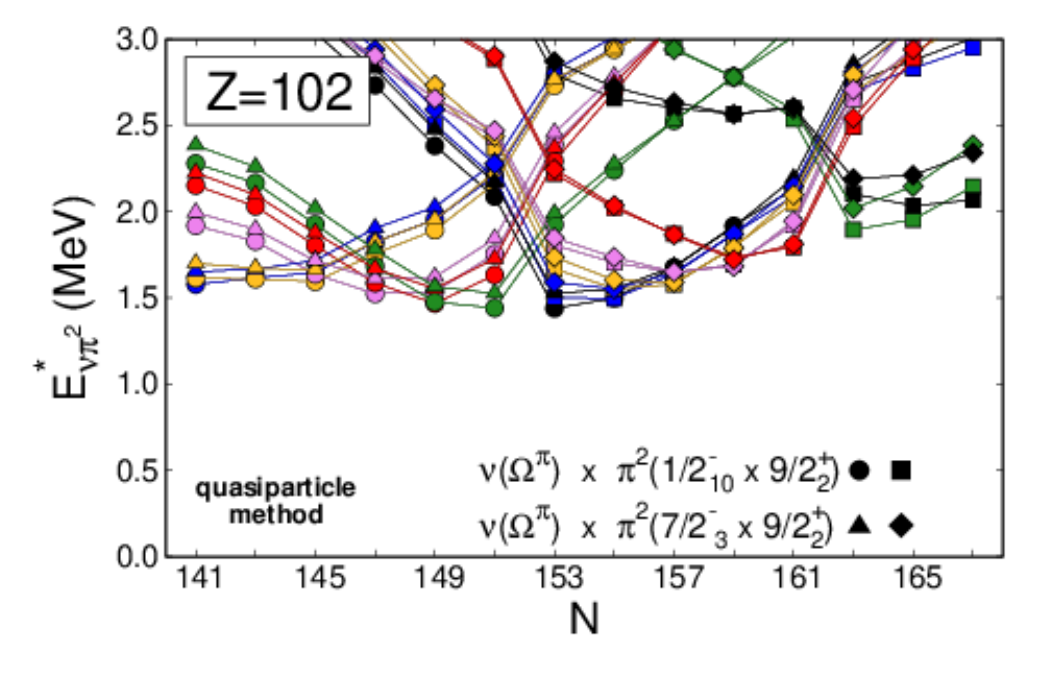}
\caption{As in Fig. \ref{1n2p_100} but for No isotopes. Configurations with the
 lowest neutron states $\Omega_{\nu}^{\pi}$ coupled to the two-proton pair
$\pi^2\{1/2^-_{\sf 10}[521] \otimes 9/2^+_{\sf 2}[624]\}$
are depicted by circles and squares with the colour-coding as in Fig.
\ref{1n2p_100}. The same neutron states coupled to the proton pair
$\pi^2\{7/2^-_{\sf 3}[514]\otimes 9/2^+_{\sf 2}[624]\}$ are marked
 by the same colours, with circles replaced by triangles and squares by
 diamonds.  }
\label{1n2p_102}
\end{figure}

\begin{figure}[!tbp]
\includegraphics[width=0.8\linewidth]{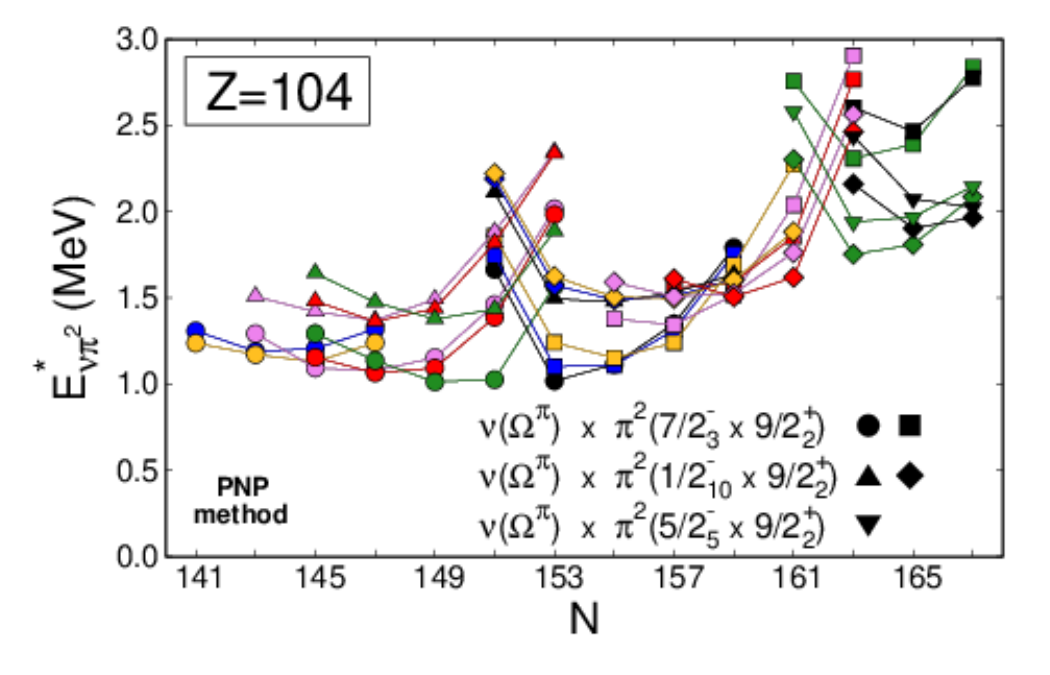}
\includegraphics[width=0.8\linewidth]{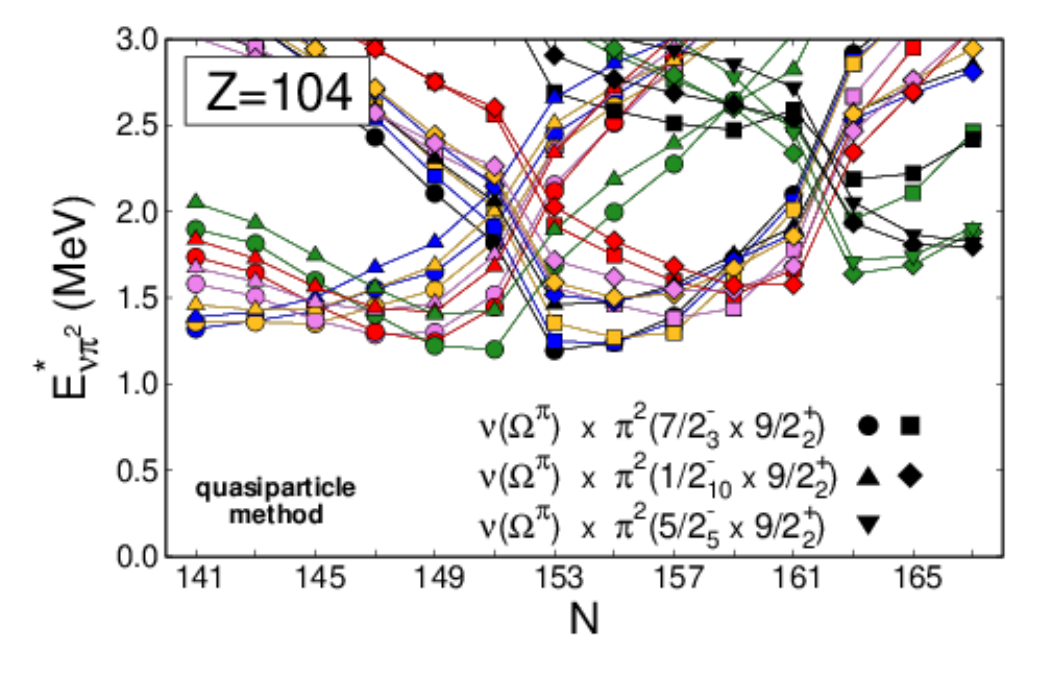}
\caption{As in Fig. \ref{1n2p_100} but for Rf isotopes. Configurations
 with two-proton pairs $\pi^2(7/2^-_{\sf 3} \otimes 9/2^+_{\sf 2})$
 and $\pi^2(1/2^-_{\sf 10} \otimes 9/2^+_{\sf 2})$ are identified
 by their neutron components with, respectively,  coloured circles/squares and
 triangles/diamonds as in Fig. \ref{1n2p_100}. Configurations with the
 pair $\pi^2(5/2^-_{\sf 5} \otimes 9/2^+_{\sf 2})$ pair are marked with
 inverted triangles, denoting the same neutron components as
  likely coloured squares and diamonds. }
\label{1n2p_104}
\end{figure}

\begin{figure}[!tbp]
\includegraphics[width=0.8\linewidth]{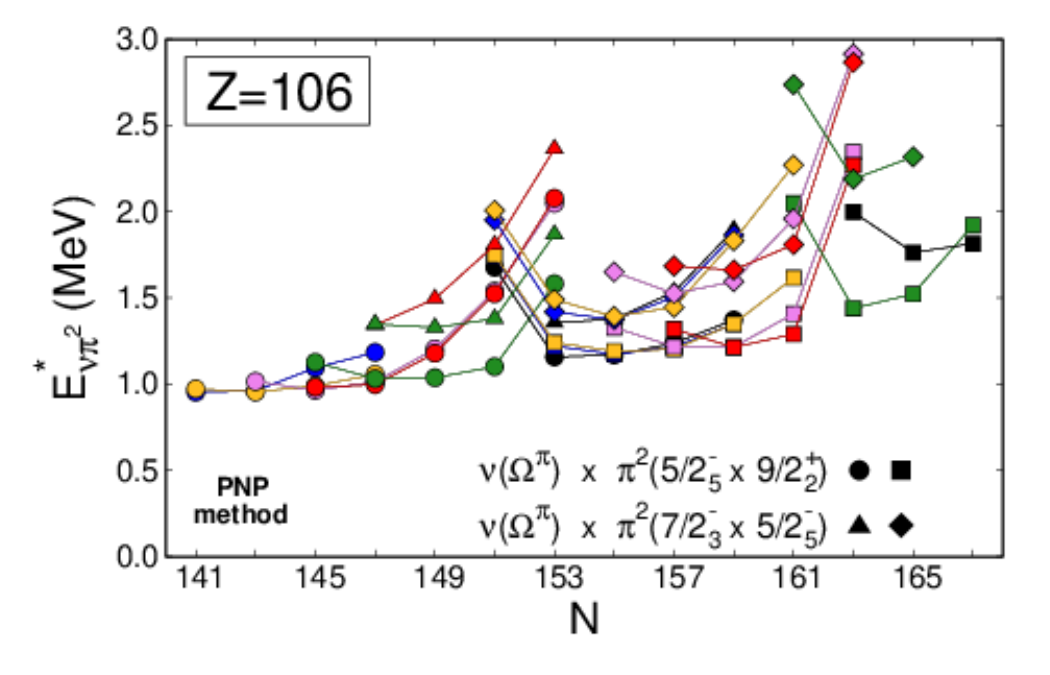}
\includegraphics[width=0.8\linewidth]{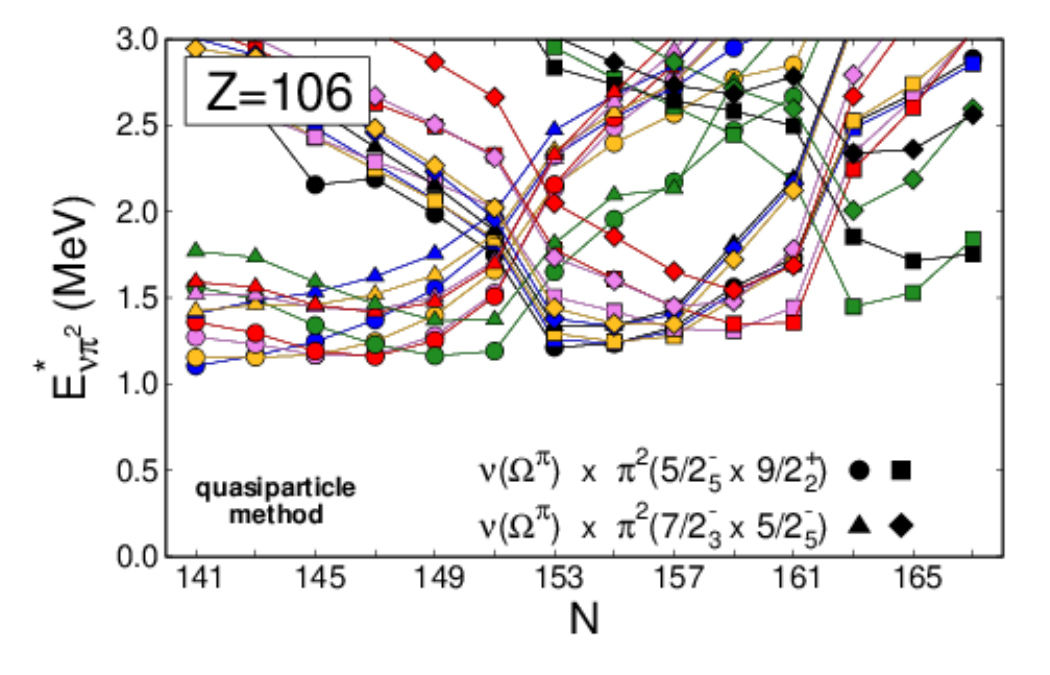}
\caption{As in Fig. \ref{1n2p_100} but for Sg isotopes. Coloured circles and
 squares refer to configurations with the $\pi^2\{5/2^-_{\sf 5}[512]
 \otimes 9/2^+_{\sf 2}[624]\}$ pair, triangles and diamonds - to those
 with the $\pi^2\{5/2^-_{\sf 5}[512]\otimes 7/2^-_{\sf 3}[514]\}$ pair,
 with coding analogous as in Fig. \ref{1n2p_100}.}
\label{1n2p_106}
\end{figure}

\begin{figure}[!tbp]
\includegraphics[width=0.8\linewidth]{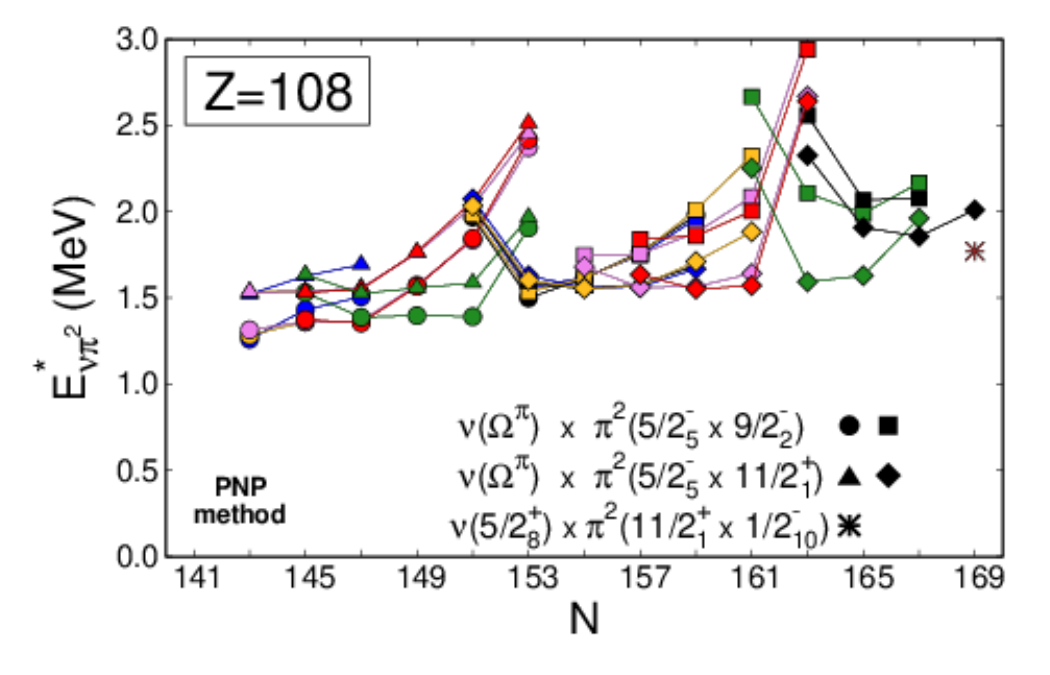}
\includegraphics[width=0.8\linewidth]{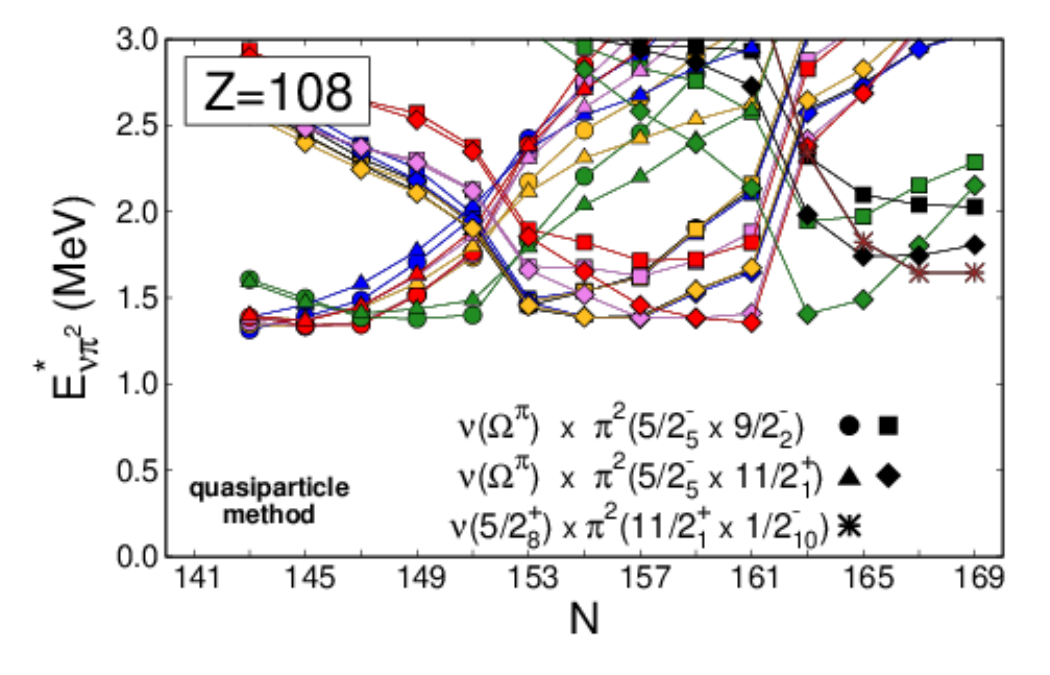}
\caption{As in Fig. \ref{1n2p_100} but for Hs isotopes. Configurations
 with the $\pi^2(5/2^-_{\sf 5} \otimes 9/2^-_{\sf 2})$ pair (circles/squares)
 and $\pi^2(5/2^-_{\sf 5} \otimes 11/2^+_{\sf 1})$ pair (triangles/diamonds)
 are marked in way analogous as in Fig. \ref{1n2p_106}. A specific
 configuration $\nu 5/2^+_{\sf 8}\otimes \pi^2\{11/2^+_{\sf 1}\otimes
 1/2^-_{\sf 10}\}$ is denoted by a brown asterisk. }
\label{1n2p_108}
\end{figure}

\begin{figure}[!tbp]
\includegraphics[width=0.8\linewidth]{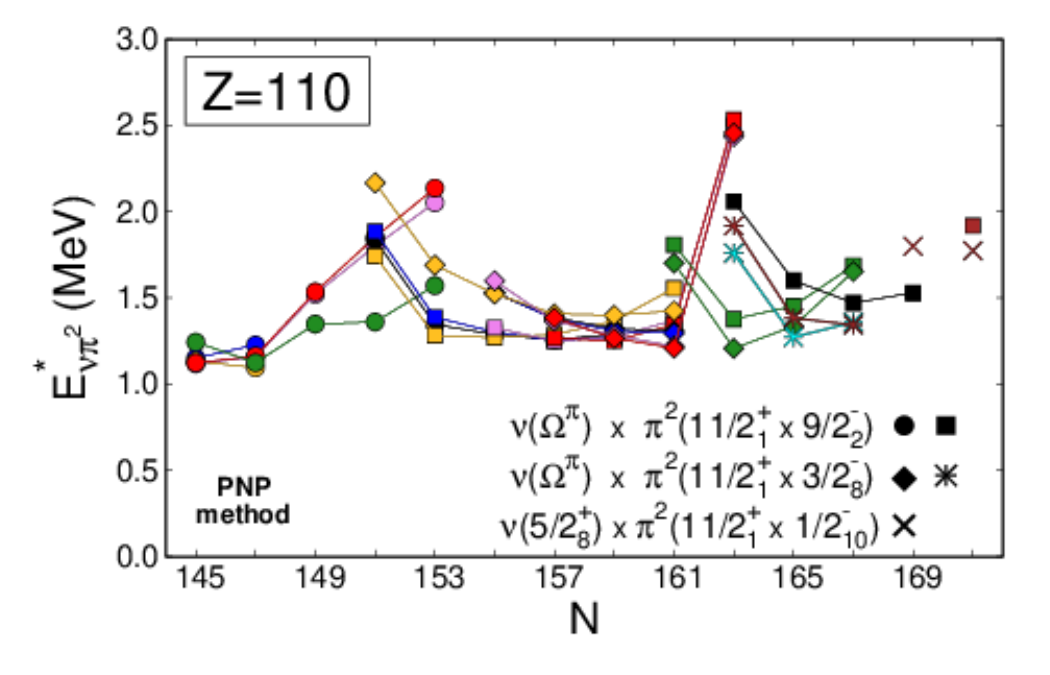}
\includegraphics[width=0.8\linewidth]{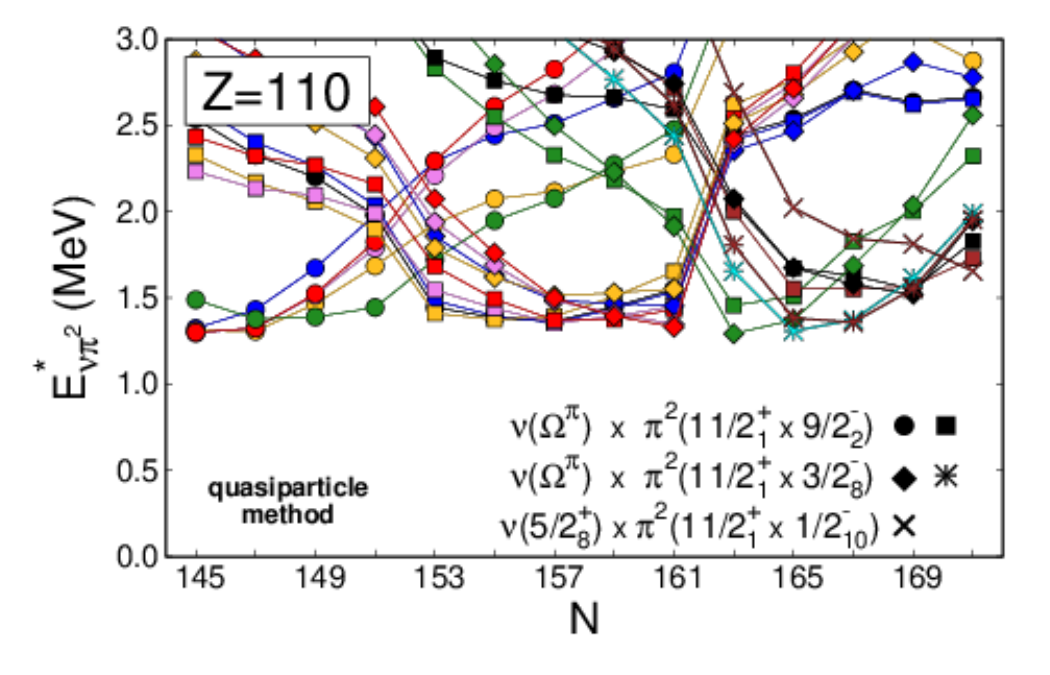}
\caption{As in Fig. \ref{1n2p_100} but for Ds isotopes. Configurations
 with the $\pi^2\{11/2^+_{\sf 1}[615]\otimes 9/2^-_{\sf 2}[505]\}$ pair are
 marked with coloured circles/squares as in Fig. \ref{1n2p_100}; the brown
square marks the same proton pair coupled to the $\nu 5/2^+_{\sf 8}[613]$
 state.
 Diamonds mark the $\pi^2\{11/2^+_{\sf 1}\otimes
 3/2^-_{\sf 8}\}$ pair coupled to the same neutron components as depicted by
 the likely coloured squares; the same
 pair coupled to the $\nu 3/2^+_{\sf 11}$ and $\nu 5/2^+_{\sf 8}$
 states are depicted by cyan and brown asterisks; the brown cross marks
 the $\nu 5/2^+_{\sf 8}\otimes \pi^2\{11/2^+_1\otimes 1/2^-_{\sf 10}\}$
 configuration.  }
\label{1n2p_110}
\end{figure}

\begin{figure}[!tbp]
\includegraphics[width=0.8\linewidth]{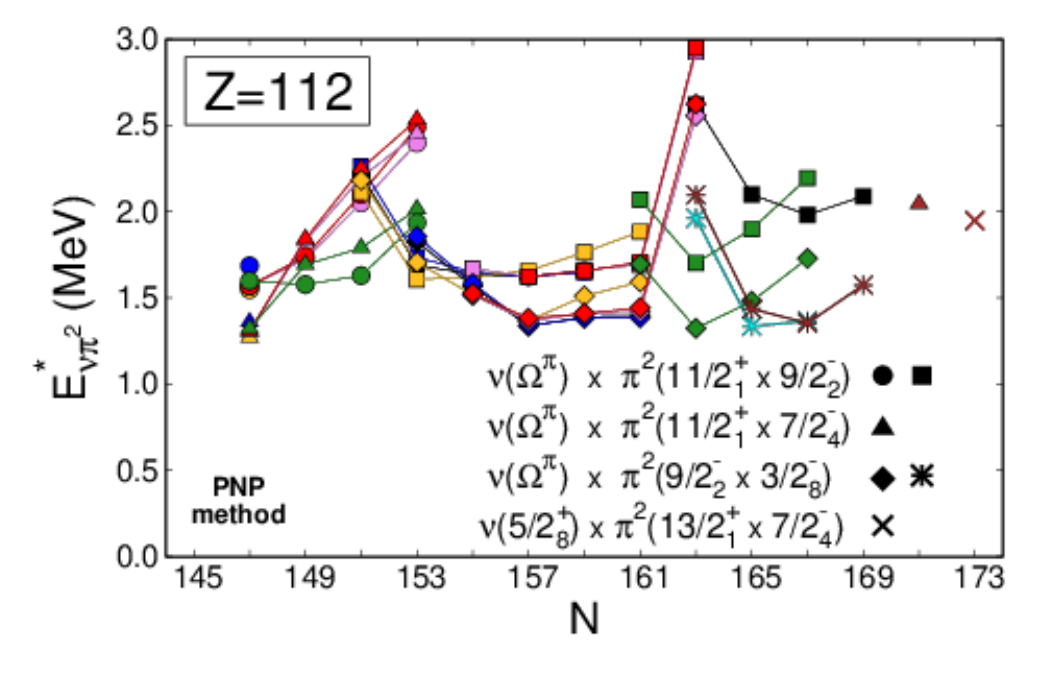}
\includegraphics[width=0.8\linewidth]{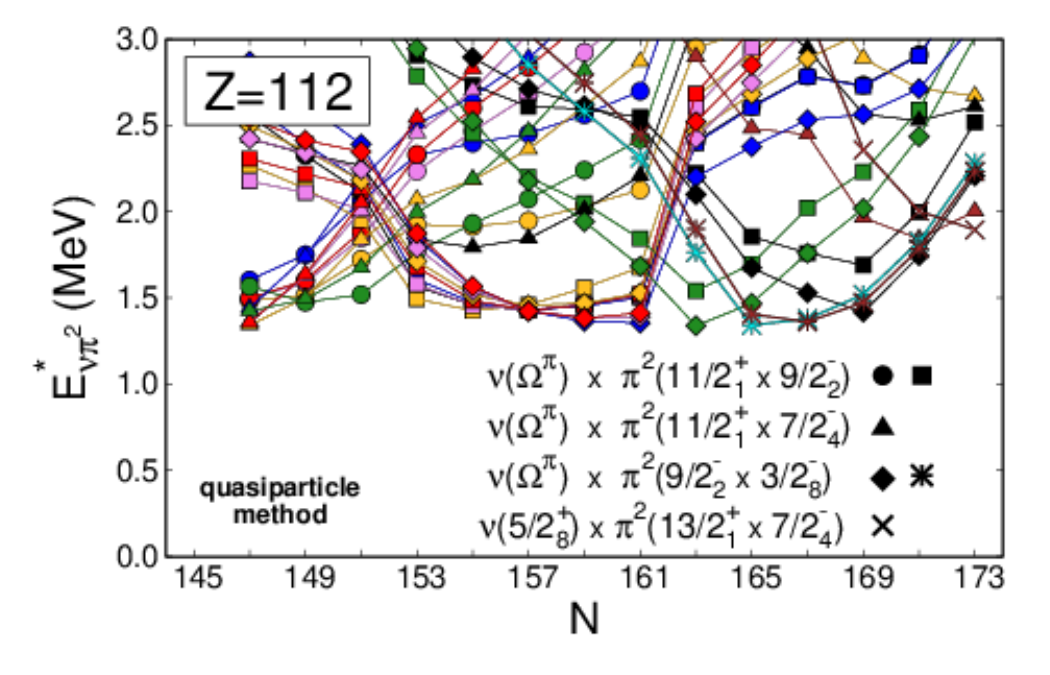}
\caption{As in Fig. \ref{1n2p_100} but for Cn isotopes. Configurations
with the $\pi^2\{11/2^+_{\sf 1}[615]\otimes 9/2^-_{\sf 2}[505]\}$ pair are
 marked by circles/squares as in Fig. \ref{1n2p_110}. Triangles depict
 the $\pi^2\{11/2^+_{\sf 1}\otimes 7/2^-_{\sf 4}\}$ pair coupled to the same
 neutron states as specified by likely coloured circles, except for the brown
 triangle which marks the same pair coupled to the $\nu 5/2^+_{\sf 8}$ state.
 Configurations with the $\pi^2\{9/2^-_{\sf 2}\otimes 3/2^-_{\sf 8}\}$ pair
 marked by diamonds and asterisks have the same neutron components as likely
 marked configurations in Fig. \ref{1n2p_110}; the brown cross marks the
 $\nu 5/2^+_{\sf 8}\otimes \pi^2\{13/2^+_{\sf 1}\otimes 7/2^-_{\sf 4}\}$
 configuration.  }
\label{1n2p_112}
\end{figure}

  The low-lying $1\nu2\pi$ high-K configurations
  involve single-neutron states situated near the neutron
 Fermi level coupled to low-lying two-proton ($2\pi$) excitations
 characterized by large $K$ values. In Figures \ref{1n2p_100}
 through \ref{1n2p_112} are presented excitation energies
 $E_{\nu\pi^2}^{*}$ of such configurations in isotopes of studied
 elements obtained from the quasiparticle method (the bottom parts)
 and the PNP calculation (the top parts). As may be seen there,
 the composition of the favoured $2\pi$ excitations changes
 with increasing proton number $Z$ from Fm to Cn. These changes can be
 also seen in Fig. \ref{plev2}, where s.p. proton levels are shown for four chosen nuclei at their calculated equilibrium deformations.

 For Fm ($Z=100$), it is the $\pi^2 7^- \{7/2^-_{\sf 3}\otimes
 7/2^+_{\sf 3}\}$ proton pair, leading to excitation energies
 $\gtrsim$ 1.5~MeV. In No ($Z=102$) and Rf ($Z=104$), configurations
 such as $\pi^2 5^- \{ 9/2^+_{\sf 2}\otimes 1/2^+_{\sf 10}\} $ and
 $\pi^2 8^- \{7/2^-_{\sf 3}\otimes 9/2^+_{\sf 2}\}$ become energetically
 favorable, with some states in Rf exhibiting excitation energies around
 1.0--1.2~MeV, particularly in the $N \approx 147$--155 region. In Sg
 ($Z=106$), proton pairs like $\pi^2 6^+ \{7/2^-_{\sf 3}\otimes
 5/2^-_{\sf 5}\}$ and $\pi^2 7^- \{9/2^+_{\sf 2}\otimes 5/2^-_{\sf 5}\}$
 contribute to low-lying states, with the quasiparticle calculations
 indicating minima around 1.1~MeV for $N \approx 145$--155. The influence
 of the $Z=108$ proton shell closure becomes apparent in Hs, where
 configurations involving orbitals above this gap, such as
 $\pi^2 7^+ \{5/2^-_{\sf 5}\otimes 9/2^-_{\sf 2}\} $ and
 $\pi^2 8^- \{5/2^-_{\sf 5}\otimes 11/2^+_{\sf 1}\}$, are significant,
 with calculated excitation energies generally above 1.3~MeV. Finally,
 for the heaviest elements considered, Ds ($Z=110$) and Cn ($Z=112$),
  key proton pairs are $\pi^2 10^- \{9/2^-_{\sf 2}\otimes
 11/2^+_{\sf 1}\}$, $\pi^2 7^- \{3/2^-_{\sf 8}\otimes 11/2^+_{\sf 1}\}$,
 and $\pi^2 6^+ \{3/2^-_{\sf 8}\otimes 9/2^-_{\sf 2}\}$, with the
 lowest quasiparticle excitation energies predicted around 1.2--1.3~MeV,
 particularly for Ds near $N=163$, while energies in Cn are generally
 somewhat higher.

 Each of the Figures \ref{1n2p_100}-\ref{1n2p_112} presents
 results for {\it the same configurations}. While
 the energies obtained for selected configurations in the quasiparticle
 method are shown for all even-odd isotopes (within the chosen energy
 window), the PNP energies were calculated mostly over isotope ranges
 where they would be among the lowest ones. Hence a more
 comprehensive view of an isotopic dependence of $E_{\nu\pi^2}^*$ for each
 configuration can be obtained from the bottom parts of these figures.

 One feature visible in Figures \ref{1n2p_100}-\ref{1n2p_112} is an abrupt
 change in energy of $1\nu2\pi$ configurations between isotopes $N-1$ and
 $N+1$ for $N=152$ and 162.
 Its origin are the gaps in the s.p. neutron spectrum at $N=152$ and $N=162$,
 as a neutron state just below the gap, close to the Fermi level at $N-1$,
 becomes distant from it at $N+1$, while a level above the gap
 at $ N-1$ becomes closer to the Fermi level at $N+1$.
 Another noticeable feature is the increase in $1\nu$–$2\pi$ energies with $N$, very pronounced in Fm, less so in No and Rf, and even weaker for larger $Z$. This behavior largely reflects the variation of the proton pairing gap $\Delta$: it increases with $N$ in Fm, No, and Rf, but the increase becomes progressively smaller with $Z$, and disappears entirely for Sg–Cn.

 Below, we comment on results for successive isotopic chains.


\textbf{\textit{Fm}}\ \
 The majority of the lowest-lying high-$K$
 1$\nu$2$\pi$ configurations in Fm isotopes contain the
 $\pi^2 7^-\{7/2^+_{\sf 3}\otimes 7/2^-_{\sf 3}\}$ pair
 coupled to the lowest-lying one-neutron states
 (still lower lying configurations with the
 $\pi^2 4^-\{7/2^+_{\sf 3}\otimes 1/2^-_{\sf 10}\}$ pair are
 omitted as having $K_{2\pi}<5$).
  Energies from the PNP formalism, shown in the upper panel of
 Fig. \ref{1n2p_100}, rise with $N$ from 1.3 to $\approx$2.8 MeV.
 The lowest energies, around 1.3--1.4~MeV, are predicted for the lightest
 isotopes ($N=141$--143). A moderate increase in energy is observed up to
 $N=149$, followed by a significant rise for heavier isotopes.
 Configurations with the proton pair
 $\pi^2 5^-\{7/2^+_{\sf 3}\otimes 3/2^-_{\sf 7}\}$ (not shown in figure
 as having smaller $K_{2\pi}$)
 lie slightly lower in $N=145,147,149$ isotopes, where they are
  nearly degenerate with those shown in Fig. \ref{1n2p_100}.


The quasiparticle method (lower panel in Fig. \ref{1n2p_100})
 yields a qualitatively similar picture for excitation energies,
 although their rise with $N$ is milder than for the PNP results.

 Overall, excitation energies obtained for even-odd Fm nuclei suggest that
 low-lying $1\nu2\pi$ high-K isomers could be expected only in
 lighter isotopes.

\textbf{\textit{No}}\ \ 
 The lowest-lying $1\nu2\pi$ high-$K$ configurations
 have two-proton components $\pi^2 5^- \{1/2^-_{\sf 10}
 \otimes 9/2^+_{\sf 2}\}$ (circles and squares in Figure~\ref{1n2p_102})
  and $\pi^2 8^- \{7/2^-_{\sf 3} \otimes 9/2^+_{\sf 2}\}$ (triangles and
 diamonds) - Fig. \ref{1n2p_102}.

 In the PNP calculations (upper panel)
 the configurations involving the $\pi^2\{1/2^-_{\sf 10} \otimes
 9/2^+_{\sf 2}\}$ pair (circles, squares) lie systematically lower in
 lighter isotopes, reaching a minimum of about 1.45~MeV for $N=149, 151$
  (green circles) and 153 (black circle). In contrast, the configurations
 built on the $\pi^2\{7/2^-_{\sf 3} \otimes 9/2^+_{\sf 2}\}$ pair
 (triangles, diamonds) are predicted at excitation energies above 1.7~MeV
 for $N \le 151$. This comes from the $K_{2\pi}=8^-$ state being non-optimal
  which generates an additional energy cost.
 The energy difference between two sets of configurations decreases towards
  $N=159$, where they become nearly degenerate.
 All energies sharply increase with $N$ for $N>152$ to $\approx 2.5$ MeV at
 $N=167$.


 In the quasiparticle method (lower panel), the separation between
 energies of configurations involving $K_{2\pi}=5$ and 8 two-proton
 pairs is small, $\le 100$ keV  (as in the quasiparticle method a downward
 shift of the Fermi level is not accounted for).
 While the lowest energy of $\approx 1.45$~MeV is also found at $N=149 -
 153$ (green and black circles), the rise in excitation energies with $N$
 for $N>152$ is less steep.

 Energies of excitation above the rotational g.s. structure,
 discussed in Sect. II, estimated in the quasiparticle method
 are: in $N=151$, for the $\nu 9/2^-_{\sf 3}$ g.s.
 configuration coupled to $K_{2\pi}=5$ and $K_{2\pi}=8$ pairs (green circle
 and trangle) - respectively $\approx 1.1$ MeV and $\approx 0.75$ MeV;
 in $N=153$, for the $\nu 1/2^+_{\sf 14}$ g.s. coupled to
  $K_{\pi}=5$ and $K_{2\pi}=8$ pairs (black circle and triangle) -
 $\approx 1.3$ MeV and $\approx 1.1$ MeV. In the PNP method,
 these excitations are similar for states involving the $K_{2\pi}=5$ pair,
 but 300 - 400 keV larger for those with the $K_{2\pi}=8$ component.
 This does not suggest an isomeric character of the state composed
 of the $8^-$ proton pair and the $1/2^+_{14}$ neutron in $^{255}$No.
 Slightly smaller excitations are estimated in the quasiparticle method
 for configurations with the $\nu 11/2^-_{\sf 2}$ state coupled to both
 proton pairs (yellow square and diamond) in $^{255}$No, but they remain
 large in the PNP method (see the comparison with the the experimental
 data at the end of this section).

In conclusion, both calculation methods point to the $N \approx 149$ region
 as having the lowest excitation energies, around 1.4--1.5~MeV.
 Since in the quasiparticle method
  energies of the higher-$K$ states involving the $8^-$ proton pair
 are lower, it gives a more favourable prediction for their isomeric
 character; within the more reliable PNP method such a conclusion is
 uncertain.

\textbf{\textit{Rf}}\ \ 
  The same two-proton pairs as in No enter the
 lowest-lying high-$K$ $1\nu2\pi$ states in Rf isotopes -
 see Fig. \ref{1n2p_104}. As the
 $K_{2\pi}=8$ pair is now the optimal one (see Fig. \ref{plev2}) the order
 of two types of configurations is inverted (note the exchange of symbols
between Figures \ref{1n2p_102} and \ref{1n2p_104}), except for $N=161$-167.
 In addition, in the heaviest isotopes, the proton pair
$\pi^2 7^-\{9/2^+_{\sf 2}\otimes 5/2^-_{\sf 5}\}$ appears as forming
 the second lowest $1\nu2\pi$ configuration in $N=167$, and,
 within the quasiparticle method, in $N=165$.

 It is noteworthy that $1\nu2\pi$ energies calculated in Rf are clearly
 lower than in Fm and No.
 The lowest energies, $\approx 1$ MeV in PNP, and $\approx 1.2$ MeV in the quasiparticle method, occur in $N=149,151,153$ isotopes.
 Since configurations containing the higher-$K_{2\pi}$ proton pair
 are now lower in energy, their excitation above the rotational g.s.
 sequence estimated in lighter isotopes (up to $\approx 157$) is
 compatible with isomerism, for example,
 $\approx$ 0.6 - 0.8 MeV in PNP/qp method for $N=153$.

  For larger $N$, the lowest excitation energies rise to $\approx 2$
 MeV at $N=167$ in the PNP, and slightly less in the q.p. method.
  Again, the energy differences between two types of
 configurations are more pronounced in the PNP variant of calculations -
   see Fig. \ref{1n2p_104}.
  The energies predicted by the quasiparticle method are
 generally slightly higher than those of the PNP in the $N<151$ region,
 but lower than them for $N>153$.

 In summary, both methods identify Rf isotopes with $N\lesssim 155$, and
 particularly those around $N=149-151$, as the most promising for low-lying
  $1\nu2\pi$ isomers, with predicted excitation energies as low as
 1.0–1.2~MeV.

 The preceding discussion concerning even-odd Fm, No, and Rf isotopes
 changes if we assume the inverted order of the proton $1/2^-_{\sf 10}$
 and $7/2^-_{\sf 3}$ levels (cf Fig. \ref{plev2}), as it is suggested  by
 the experimentally established spins of low-lying states
  in Md and Lr isotopes around $N=150$ \cite{Kondev2015,David2015,Khuyagbaatar2020}.
 In such a case, the excitation of the proton $K_{2\pi}=7$ pair in Fm
 isotopes would be smaller and the isomeric character of configurations
 including it more probable in lighter isotopes. In No, the inversion of
 these proton levels would make the $K_{2\pi}=8$ proton pair optimal and
 would lower energies of high-$K$ congfigurations involving it. The
 consequence for Rf isotopes would be an increase in energies of
 configurations including the $K_{2\pi}=8$ proton pair and a decrease in
 those with the $K_{2\pi}=5$ pair.
 Overall, prospects for high-$K$ isomers in these nuclei would become
  more similar (this means, better in Fm and No), especially in lighter
  isotopes.


\textbf{\textit{Sg}}\ \ 
 Both in the PNP and qiasiparticle calculations, the energetically favoured
 configurations contain the $\pi^2 7^-\{5/2^-_{\sf 5}\otimes
 9/2^+_{\sf 2}\}$ pair (circles and squares in Fig \ref{1n2p_106})
 while the second lowest often involve the $\pi^2 6^+\{5/2^-_{\sf 5}\otimes
  7/2^-_{\sf 3}\}$ pair (triangles and diamonds), see also Fig. \ref{plev2}.
  Energies of the $1\nu2\pi$ high-$K$ states are the lowest for the
 lightest isotopes. In the PNP results, energies very slowly rise from
 $\approx 0.95$ MeV at $N=141$, attaining values $\approx 1.1 - 1.2$
  MeV at $N=151, 153$ and $\approx 1.25$ MeV at $N$=159 (in the
 range of known isotopes); only for $N>161$ they rise more, to $\approx
 1.8$ MeV at $N=167$.
 The quasiparticle method (lower panel) gives qualitatively similar
 results, with slightly higher lowest energies for $N \le 151$
 isotopes, reaching minimum around 1.1~MeV.


  Low energies obtained for configurations involving the $K_{2\pi}=7$ pair
  up to $N$=159 translate to relatively low excitation above the g.s.
 rotational sequence and suggest their likely isomeric character.
 This makes Sg a promising element for a search of low-lying $1\nu2\pi$
 isomers.

\textbf{\textit{Hs}}\ \ 
 Energies of the lowest high-$K$ $1\nu2\pi$ configurations in
 Hs isotopes are larger than in Sg (except for the heaviest isotopes) as
 the 2p excitations involve the promotion of one proton above the $Z$=108
 gap, see Fig. \ref{plev2}.
 The $1\nu2\pi$ states which contain the proton pair
 $\pi^2 7^+\{5/2^-_{\sf 5}\otimes 9/2^-_{\sf 2}\}$ are the
 lowest up to $N$=151 - 153 (Fig. \ref{1n2p_108}, circles) while those
 with $\pi^2 8^-\{5/2^-_{\sf 5}\otimes 11/2^+_{\sf 2}\}$ (diamonds)
 become the lowest from $N=153$ on.
 The lowest $1\nu2\pi$ configuration for $N$=169 is
 $\nu 5/8^+_{\sf 8}\otimes \pi^2 6^-\{11/2^+_{\sf 1}\otimes
 1/2^-_{\sf 10}\}$, at smaller deformation $\beta_{20}\approx 0.17$
 and $E^*=1.77$ MeV within PNP and $~100$ keV less in the q.p. method.

  In the PNP formalism (upper panel), the lowest-lying states have energies
  $1.25-1.4$~MeV for lighter isotopes ($N \le 151$), and $1.5 - 1.6$~MeV
  for $N=153 - 165$.
 The estimate of excitation above the g.s. rotational band in
 experimentally relevant isotopes $N\ge 153$ gives
 $\approx$ 0.7 and 1.1 MeV with the assumed $K_{\nu}=11/2$ and 1/2,
 respectively, so some of these configurations could be isomeric.

 In the quasiparticle calculations (the lower panel) there is a smaller
 difference between two types of configurations; between $N=141$ and 163
 they remain in the range $1.3-1.45$ MeV. Since for $N\ge 153$ these
 energies are smaller than in the PNP method, and they provide even better
 case for isomerism.

 While high-K isomers in Hs are possible, the higher excitation
 energies make them less favoured than in Rf and Sg.

\textbf{\textit{Ds}}\ \ 
  In this isotopic chain, energies of the lowest-lying high-$K$ $1\nu2\pi$ configurations are predicted smaller than for Hs.
 The lowest-energy proton-pair excitations are formed from orbitals situated
 above the $Z=108$ shell gap, see Fig. \ref{plev2}.
 The energetically favoured $1\nu2\pi$ high-$K$ states contain the
 $\pi^2 10^-\{11/2^+_{\sf 1}\otimes 9/2^-_{\sf 2}\}$ pair for $N\le 159$,
 Fig \ref{1n2p_110} (circles, squares),
 and $\pi^2 7^-\{11/2^+_{\sf 1}\otimes 3/2^-_{\sf 8}\}$ for $N\ge 161$
 (diamonds, asterisks).
  Considering isotopes $N\ge 153$ (as lighter ones are experimentally
 inaccessible), the lowest PNP energies, $E_{\nu\pi^2}^*\approx 1.3$ MeV,
 are predicted around $N=163$, see Fig. \ref{1n2p_110}, upper panel.
 Especially configurations involving
 the $K_{\pi}=10$ proton pair are good candidate for isomers (it is likely
 that the same two-proton pair is involved in the known isomer
 in $^{270}$Ds).

The quasiparticle method (lower panel) also identifies a region of low
 energies around $N=157$, with a minimum of about 1.25 MeV.

  In PNP calculations for the $N$=169 isotope, $1\nu2\pi$ configurations
  involving one of the following neutron levels:
  $11/2^+_{\sf 2}$, $5/2^+_{\sf 8}$ and $9/2^+_{\sf 4}$, all have similar
 energies, dependent primarily on the proton pair, with the lowest one for
 the $K_{\pi}=10$ pair (Fig.~\ref{1n2p_110}, black square). A similar situation occurs in the 
$N=171$ isotope, where the lowest-lying configurations are built from one of 
the neutron levels $5/2^+_{\sf 8}$, $3/2^+_{\sf 11}$, and $15/2^-_{\sf 1}$, 
coupled to the 
$\pi^2 6^-\{11/2^+_{\sf 1}\otimes 1/2^-_{\sf 10}\}$ pair, with excitation 
energies of $E^*\approx 1.8$~MeV (Fig.~\ref{1n2p_110}, brown cross). 
The energy of configurations involving the $K_{\pi}=10$ proton pair is 
$\approx 150$~keV higher when coupled to lower-$\Omega_{\nu}$ neutron states, 
and about $\approx 350$~keV higher when coupled to the neutron 
$15/2^-_{\sf 1}$ state.  

In summary, both approaches suggest that Ds isotopes are promising candidates 
for $1\nu2\pi$ isomerism in the range $N=155$--$161$, and somewhat less so 
(due to the smaller aligned $K_{2\pi}$ value) for $N=163$.

\textbf{\textit{Cn}}\ \ 
 It can be seen in Figure~\ref{1n2p_112} that
  in isotopes with $153\le N \le 169$ the lowest-lying configurations
 following from the PNP calculations
 contain the $\pi^2 6^+\{9/2^-_{\sf 2}\otimes 3/2^-_{\sf 8}\}$ pair
 (diamonds and asterisks) while those with the $\pi^2 10^-\{11/2^+_{\sf 1}
 \otimes 9/2^-_{\sf 2}\}$ pair lie $\approx$ 200-300 keV higher (very light
  isotopes probably are outside experimental possibilities).
 The lowest energies of $1\nu2\pi$ states in the PNP calculation,
  $E^*\approx 1.35$ MeV, occur around $N$=163 (upper panel). Similar
  energies in the same range of isotopes are obtained in the quasiparticle
  calculations (lower panel). However, configurations involving two proton
  pairs are closer in energy in isotopes $N=157$ - 163.
 The isomeric character of these states, although possible, is not certain as
 the $K_{2\pi}=6$ of the aligned proton pair is not large as for their
 excitation energy.

 In PNP calculations, the lowest energies of $1\nu2\pi$ configurations in the
 $N$=171,173 isotopes are around 2 MeV. The involved proton pairs are
 $\pi^2 9^-\{11/2^+_{\sf 1}\otimes 7/2^-_{\sf 4}\}$ in $N$=171 and
 $\pi^2 10^-\{13/2^+_{\sf 1}\otimes 7/2^-_{\sf 4}\}$ in $N$=173, which form
 configurations with similar energies when coupled to one of the neutrons
 states: $5/2^+_{\sf 8}$, $3/2^+_{\sf 11}$, $15/2^-_{\sf 1}$
 (and $1/2^+_{\sf 15}$ in $N$=173). Energies from the qp calculations for
 $N=171$ are smaller. In deciding whether these or the
 configurations in $N$=171, 173 isotopes of Ds could be isomeric, one has to
 take into account their reduced $\beta_{20}$ deformations of
 $\approx$ 0.13 - 0.15 ($N$=171) and 0.08 - 0.1 ($N$=173). They imply
 moments of inertia substantially smaller than in lighter isotopes and,
 consequently, smaller excitation of the 3qp states above the g.s.
 rotational band.
 Therefore, in spite of relatively high energy, configurations with large
 $K_{2\pi}$ and $K_{\nu}$ may qualify as candidates for isomers.

 In summary, for Cn isotopes, the prospects for finding low-lying $1\nu2\pi$
 isomers exist, but are weaker than in Rf and Sg due to higher energies and
 and smaller aligned angular momentum $K_{2\pi}$ of the proton pair.

\begin{figure}[!htbp]
\centering
\includegraphics[scale=0.45]{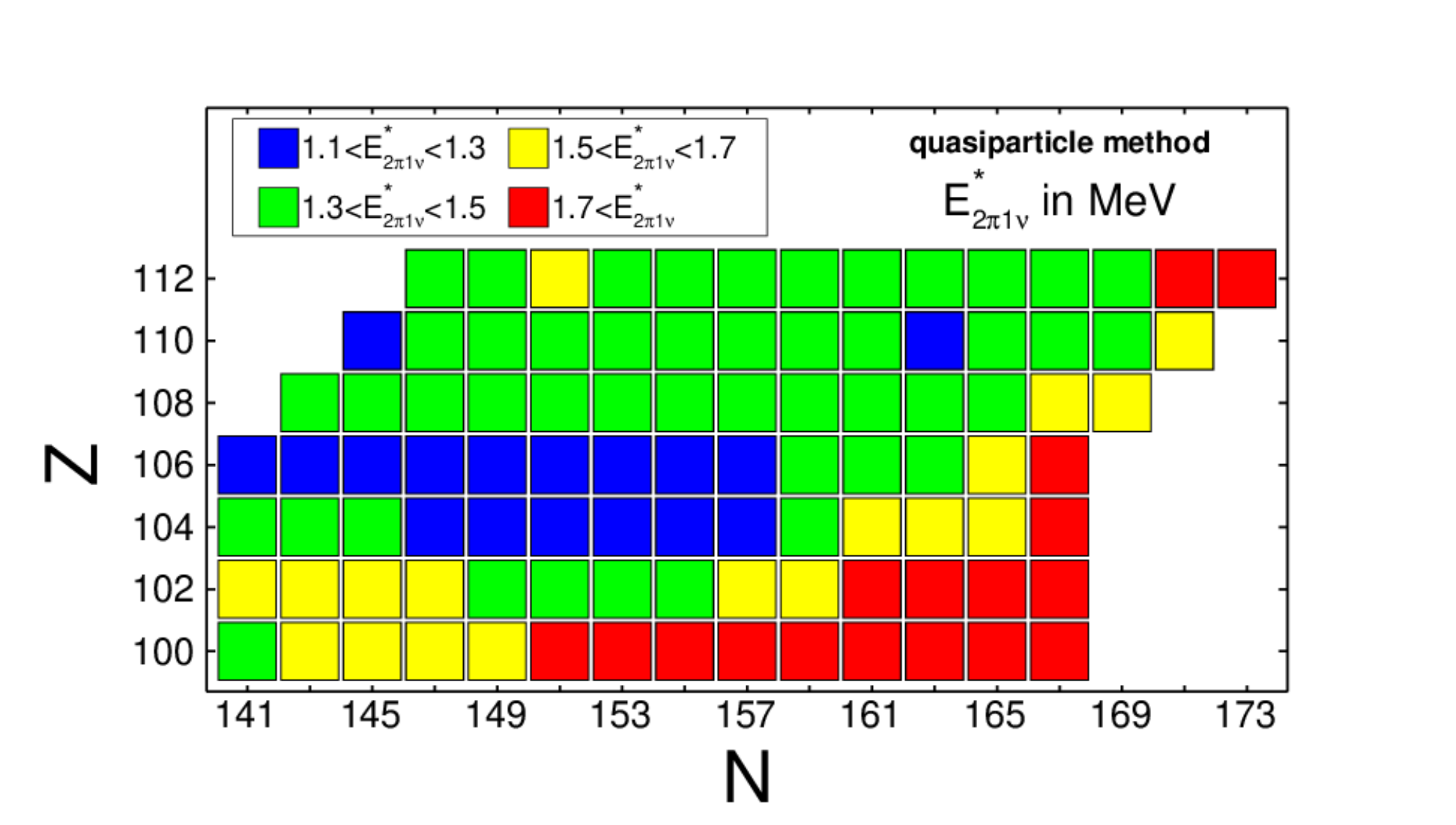}
\caption{Map of the lowest $1\nu2\pi$ 3-q.p. excitation energies calculated with the quasiparticle method.}
\label{szachnq}
\end{figure}

The calculated excitation energies for the lowest $1\nu2\pi$ quasiparticle
(q.p.) configurations, obtained using the quasiparticle method, are
 summarized in Figure~\ref{szachnq}. For the lighter elements in this
 study, specifically $Z=100$ (Fm) and $Z=102$ (No), the excitation energies
 are generally high, predominantly falling into the red ($E^* > 1.7$~MeV)
 and yellow ($1.5 < E^* < 1.7$~MeV) regions, and tend to increase with
 neutron number across the displayed range. In stark contrast, for $Z=104$
 (Rf) and $Z=106$ (Sg), a significant region characterized by lower
 excitation energies (blue, $1.1 < E^* < 1.3$~MeV) is predicted across a
 broad range of neutron numbers, approximately $N=145$--157. Outside this
 blue region, energies for Rf and Sg generally increase, transitioning
 through green ($1.3 < E^* < 1.5$~MeV) to yellow and red at higher $N$
 values. For the heavier elements investigated, $Z=108$ (Hs) and $Z=112$
 (Cn), the excitation energies are mostly found within the green range
 (1.3--1.5~MeV) and exhibit less pronounced systematic variation with $N$.
 It is noteworthy that for $Z=110$ (Ds), while the energies are generally
 in the green region, a localized minimum (blue, 1.1--1.3~MeV) appears
 around $N=163$. Consequently, the lowest excitation energies for these
 $1\nu2\pi$ configurations, as calculated with the quasiparticle method,
 are predominantly found for Rf and Sg in the $N \approx 145$--157 region,
 and for Ds in the vicinity of $N \approx 163$.


\subsection{Excitation energies of 3-neutron q.p. high-$K$ - candidates for
 isomeric states}

\begin{figure}[!htbp]
\centering
\includegraphics[width=0.9\linewidth]{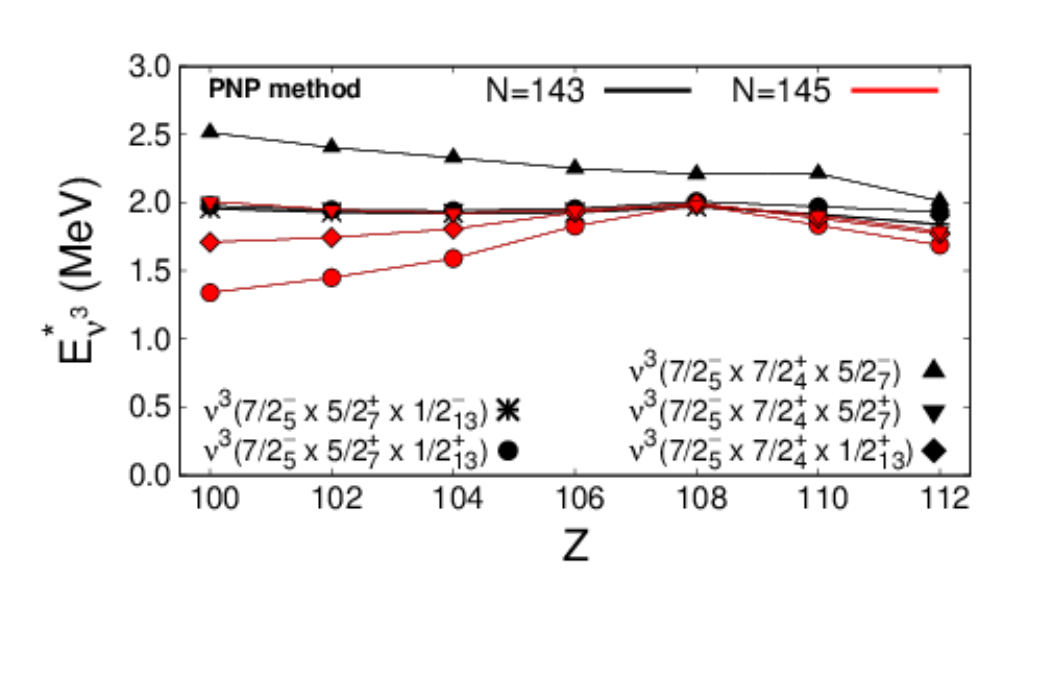}\\
\vspace{-15mm}
\includegraphics[width=0.9\linewidth]{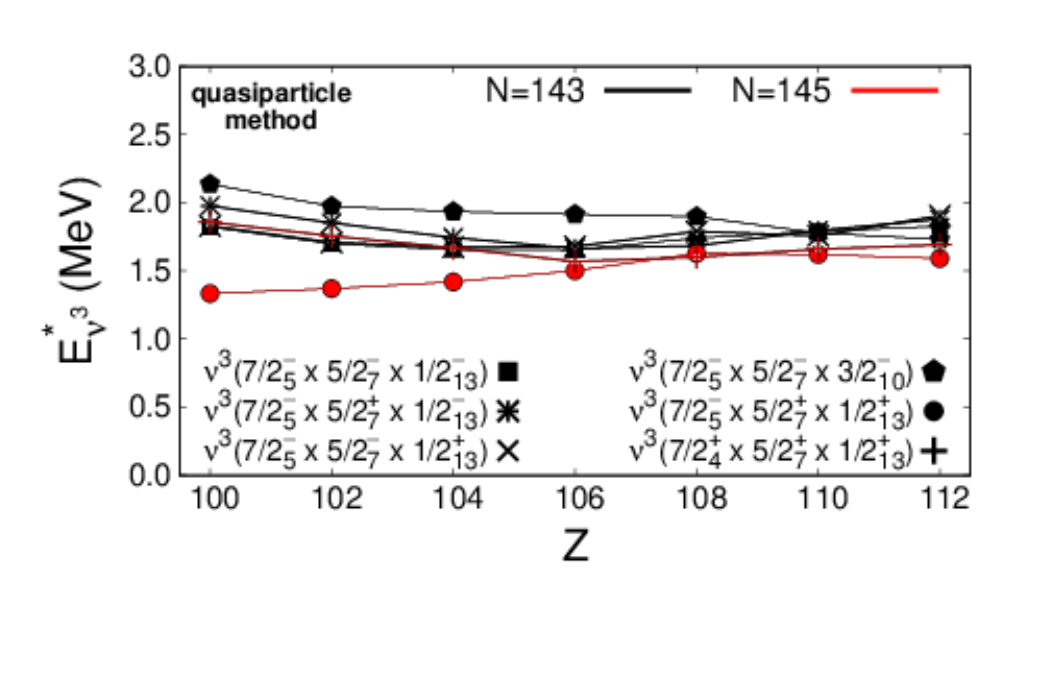}
\vspace{-15mm}
\caption{Excitation energies of low-lying high-$K$ $3\nu$
 configurations in $N=143,145$ isotones.
 Colours distinguish $N$, point shapes mark configurations.}
\label{3n15143}
\end{figure}

\begin{figure}[!htbp]
\centering
\includegraphics[width=0.9\linewidth]{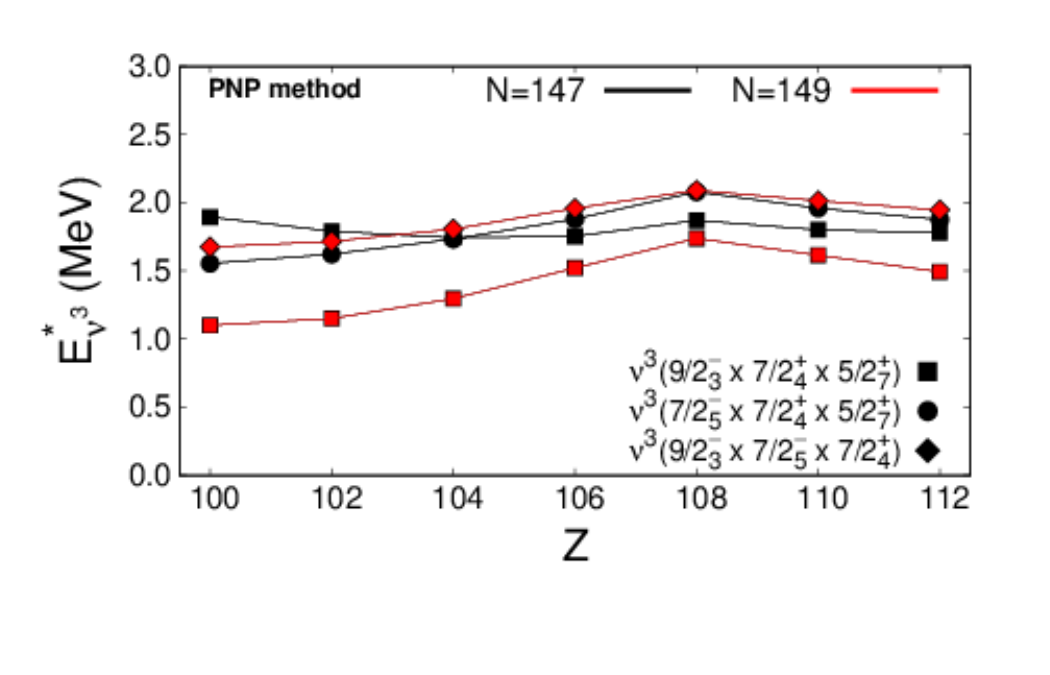}\\
\vspace{-15mm}
\includegraphics[width=0.9\linewidth]{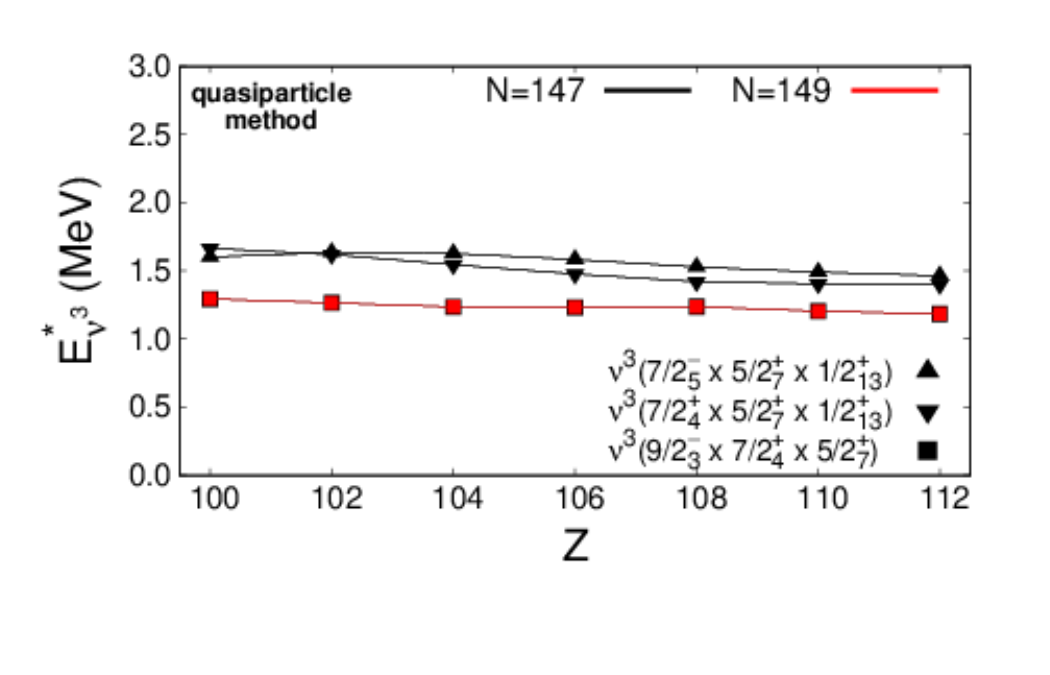}
\vspace{-15mm}
\caption{As in Fig.~\ref{3n15143}, but for
  $N=147,149$ isotones.}
\label{3n15147}
\end{figure}

\begin{figure}[!htbp]
\centering
\includegraphics[width=0.9\linewidth]{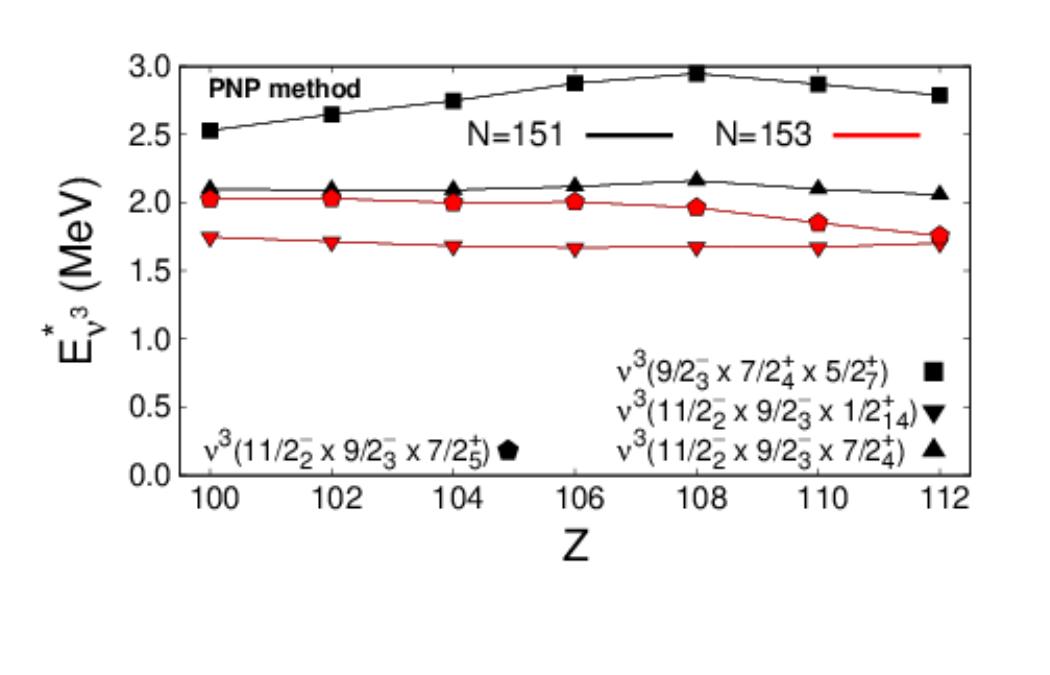}\\
\vspace{-15mm}
\includegraphics[width=0.9\linewidth]{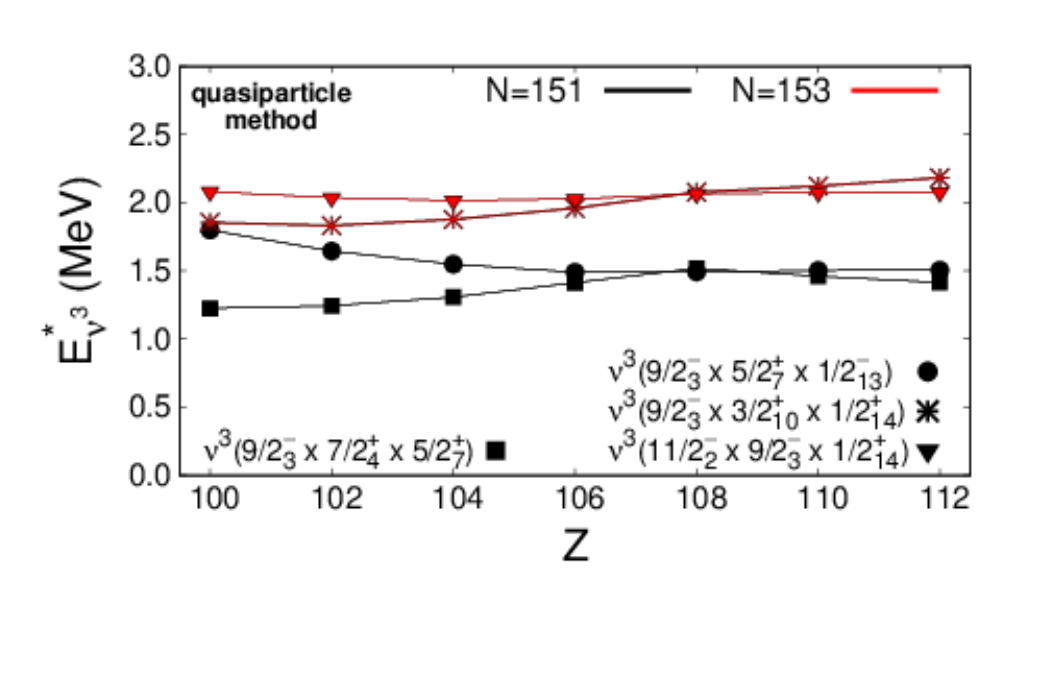}
\vspace{-15mm}
\caption{As in Fig.~\ref{3n15143}, but for
  $N=151,153$ isotones.}
\label{3n15151}
\end{figure}

\begin{figure}[!htbp]
\centering
\includegraphics[width=0.9\linewidth]{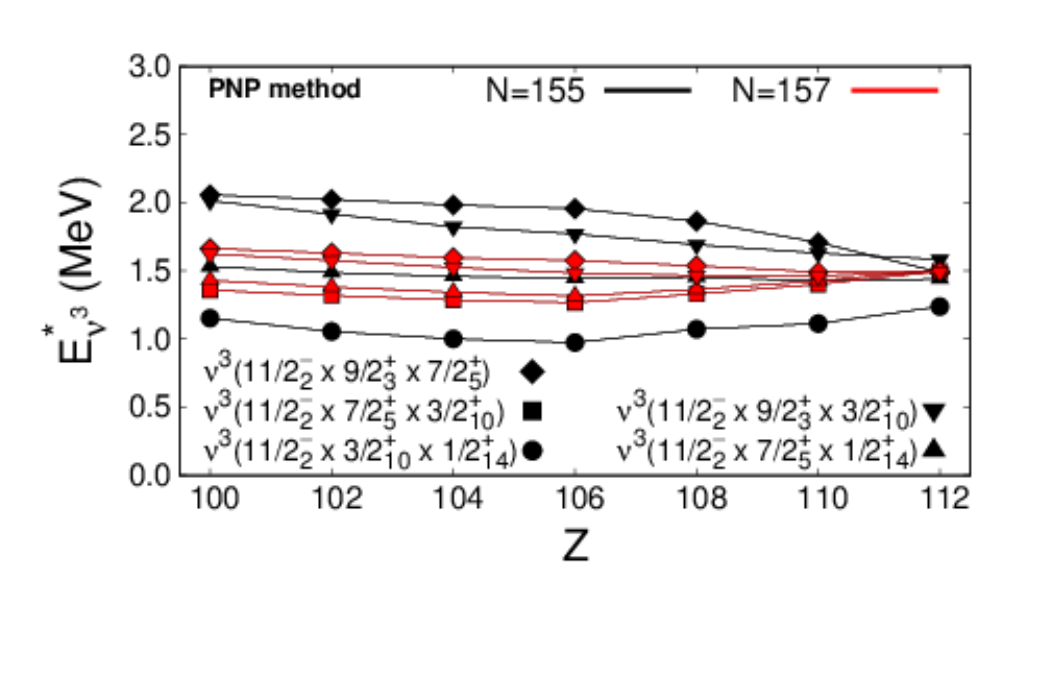}\\
\vspace{-15mm}
\includegraphics[width=0.9\linewidth]{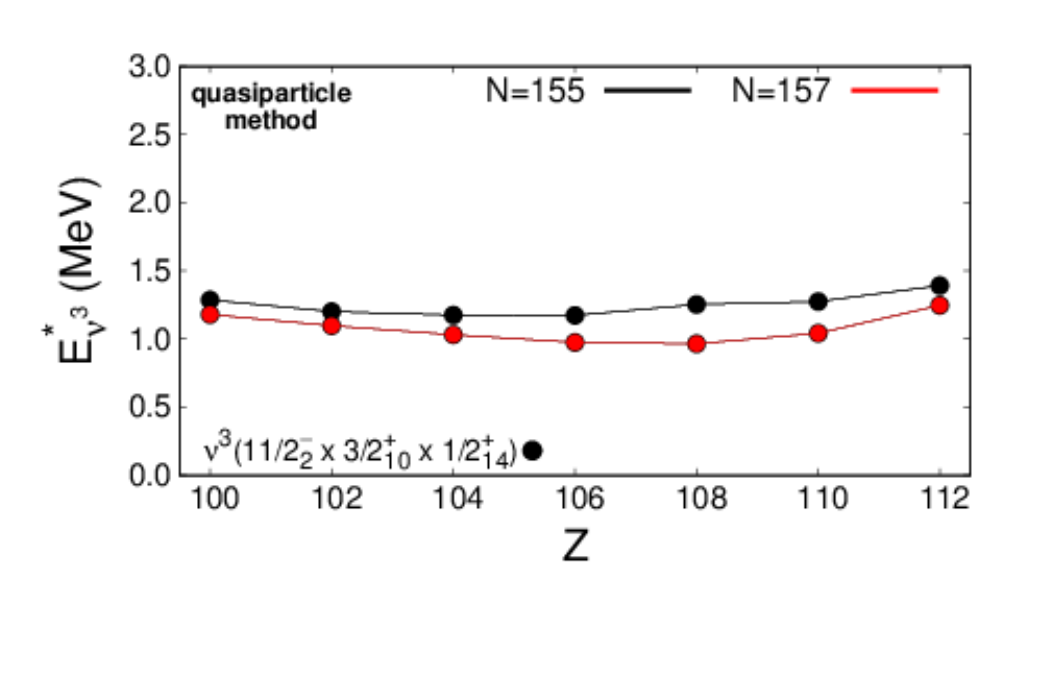}
\vspace{-15mm}
\caption{As in Fig.~\ref{3n15143}, but for
  $N=155,157$ isotones.}
\label{3n15155}
\end{figure}

\begin{figure}[!htbp]
\centering
\includegraphics[width=0.9\linewidth]{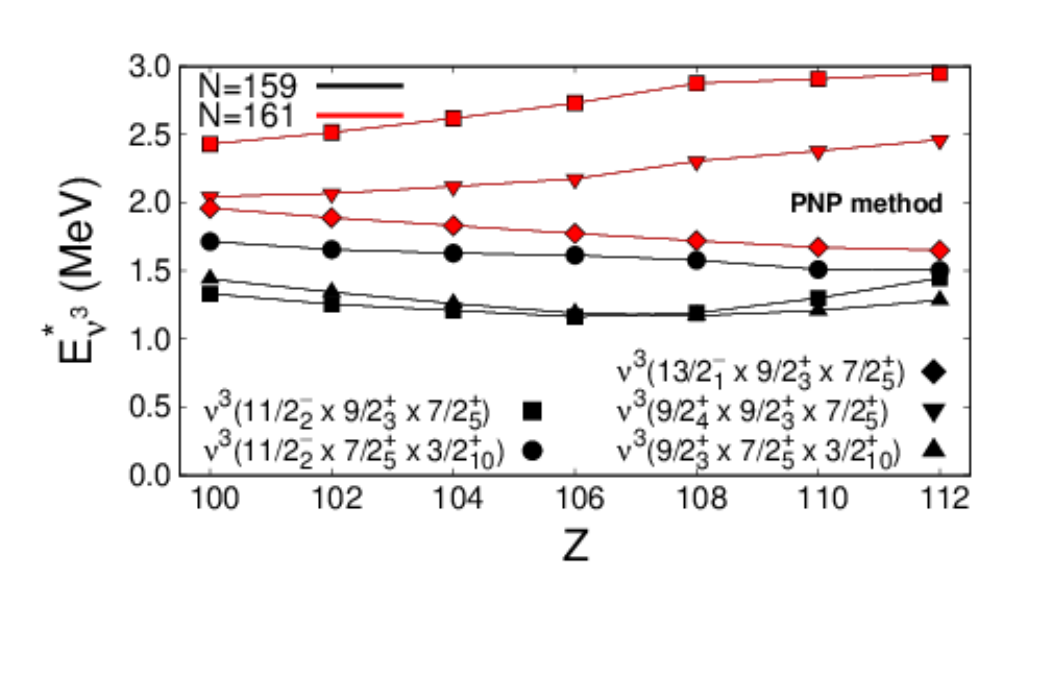}\\
\vspace{-15mm}
\includegraphics[width=0.9\linewidth]{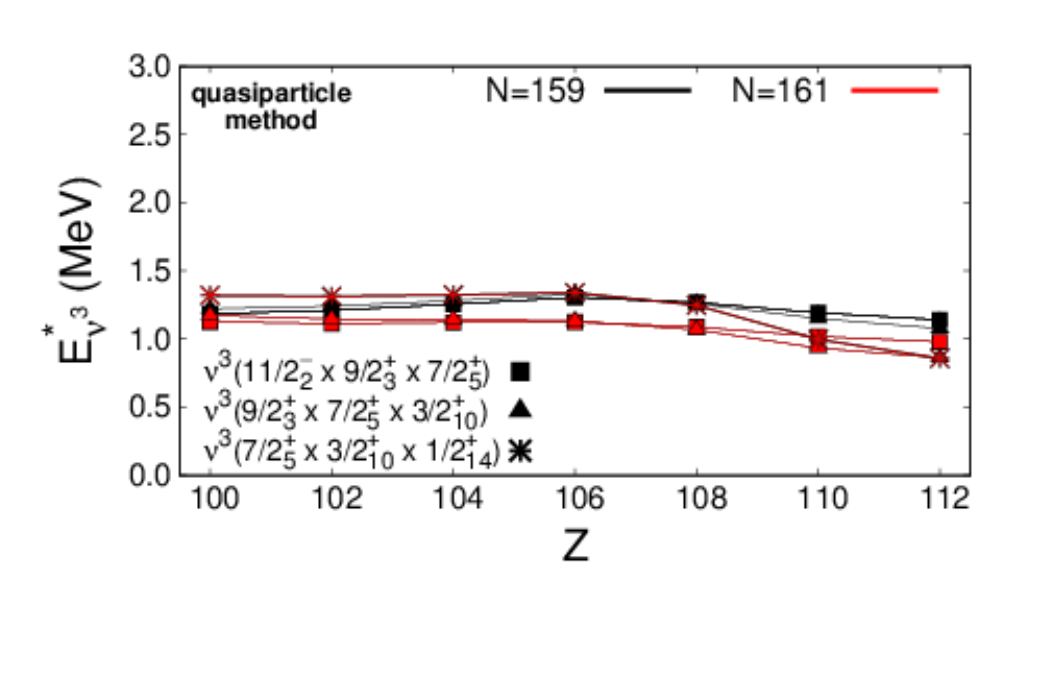}
\vspace{-15mm}
\caption{As in Fig.~\ref{3n15143}, but for
 $N=159,161$ isotones.}
\label{3n15159}
\end{figure}

\begin{figure}[!htbp]
\centering
\includegraphics[width=0.9\linewidth]{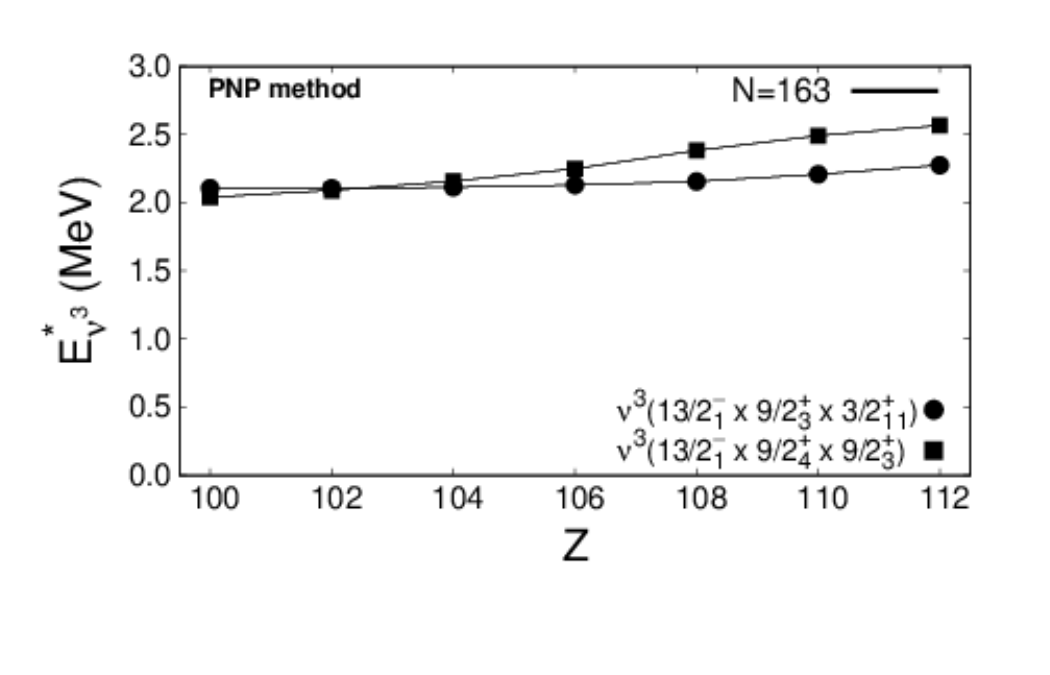}\\
\vspace{-15mm}
\includegraphics[width=0.9\linewidth]{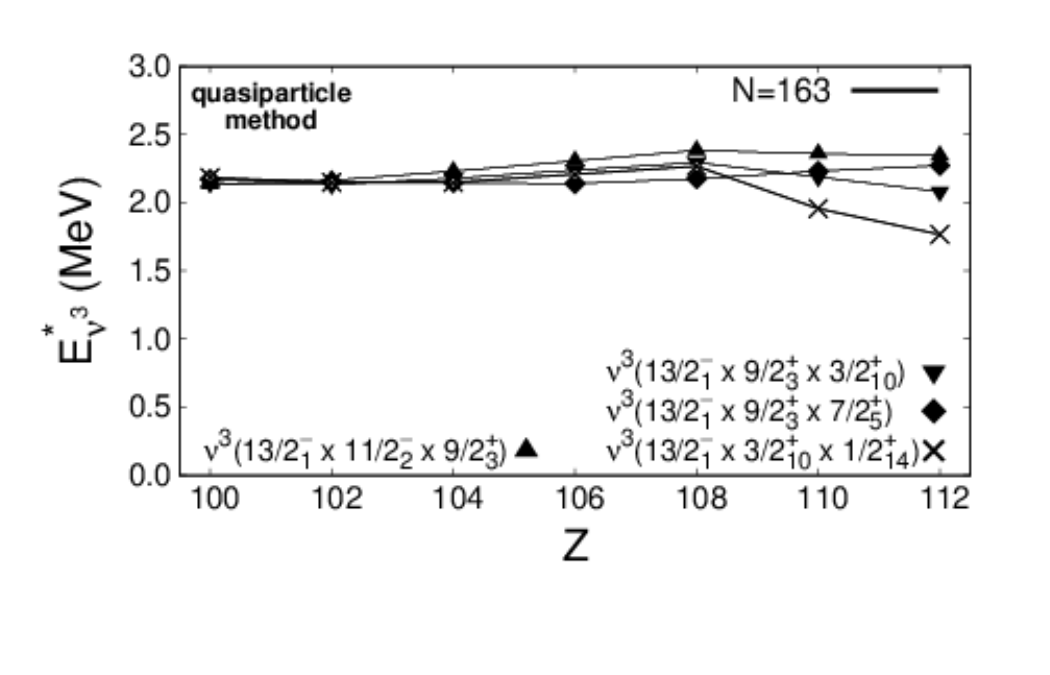}
\vspace{-15mm}
\caption{As in Fig.~\ref{3n15143}, but for
 $N=163$ isotones.}
\label{3n15163}
\end{figure}

\begin{figure}[!htbp]
\centering
\includegraphics[width=0.9\linewidth]{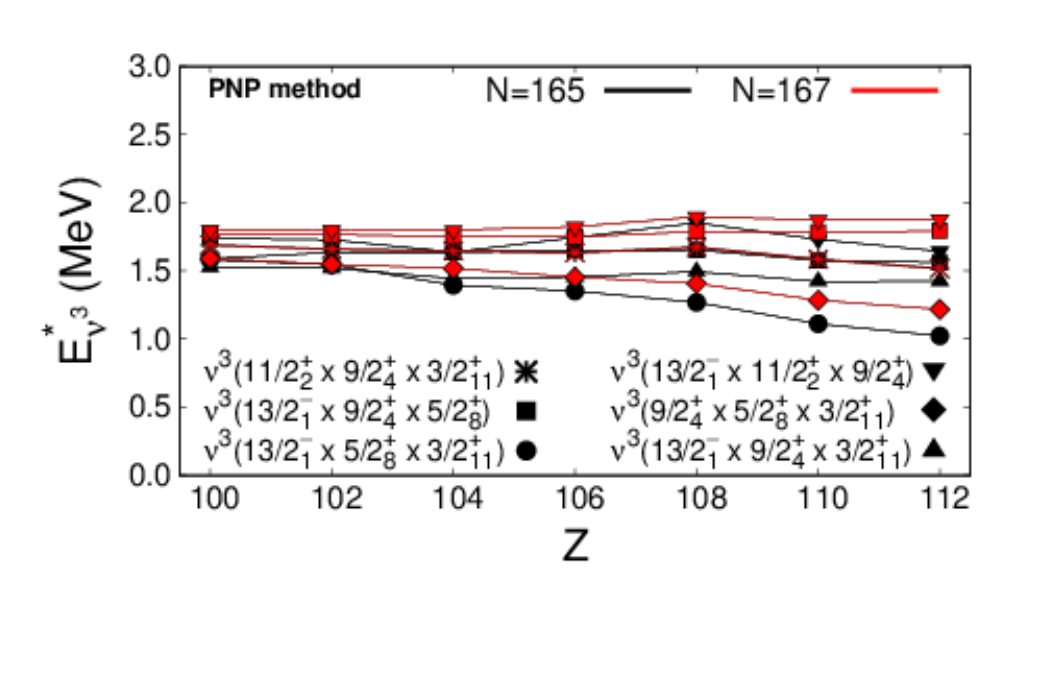}\\
\vspace{-15mm}
\includegraphics[width=0.9\linewidth]{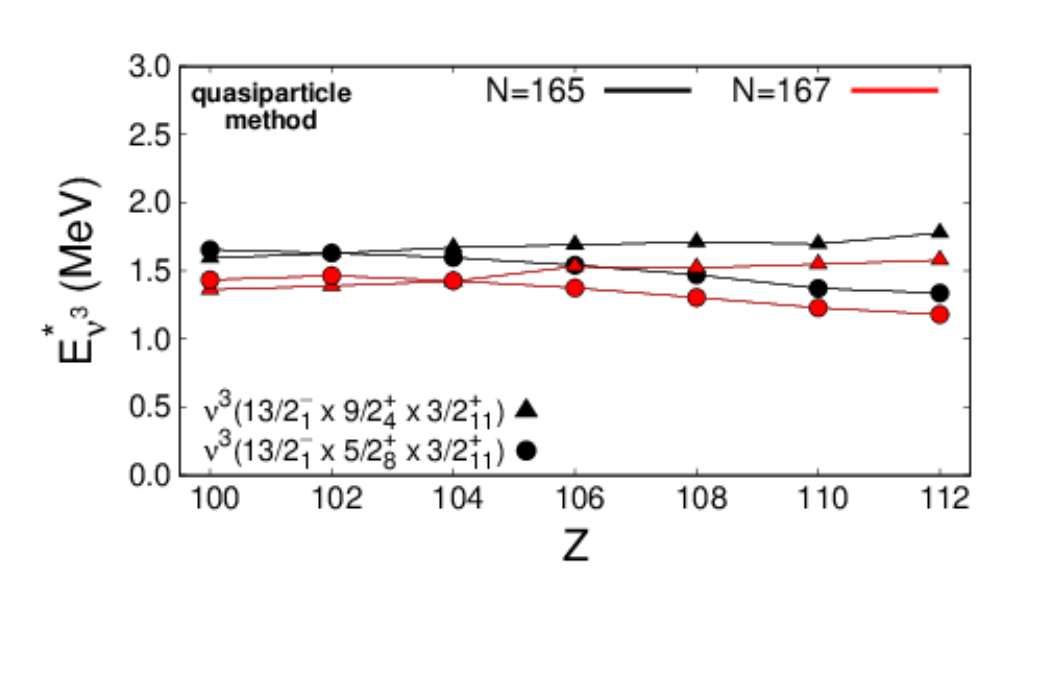}
\vspace{-15mm}
\caption{As in Fig.~\ref{3n15143}, but for
 $N=165,167$ isotones.}
\label{3n15165}
\end{figure}

 Energies for selected, mostly low-lying 3-neutron q.p. configurations
 calculated within the PNP and quasiparticle methods are shown in Figs.
 \ref{3n15143}-\ref{3n15165}.
 As mentioned in the previous section, the neutron pairing strengths
 fixed at the ratio $G_n(N,Z)/G_n^{mod}(N,Z)=\ 1.1$, with $G_n^{mod}(N,Z)$  the pairing strengths of our MM model, produced seemingly too
 low $3\nu$ excitation energies in some nuclei. Therefore, we discuss
  results of the PNP calculations with stronger pairing, with the above
  ratios fixed at 1.15.

 In each of these two methods, the lowest $3\nu$ states were independently
 selected since (as already mentioned) differences in $3\nu$
 excitation energies are expected between the q.p. and PNP results,
 especially for $N=151, 153$ and $N=161,163$ around the large gaps in the neutron s.p. spectrum.
 Results of the q.p. method in Figs. \ref{3n15143}-\ref{3n15165} were obtained
 by taking the even-even core with the neutron number $N-1$. In comparison with
 the other version, with the core of $N$ neutrons, its results are
 closer to those of PNP for $N=N_{gap}+1$, while they remain
 unphysical for $N_{gap}-1$.

 Energies $E_{\nu^3}^*$ are presented in Figures \ref{3n15143}-\ref{3n15165}
 for isotonic sequences vs $Z$. Those of the q.p method (in lower
 panels) are shown for $3\nu$ configurations that come out the lowest in some
 isotone.
 The PNP results (in upper panels) are also shown for isomer candidates
 that are lying somewhat higher, with a favourable relation of energy to $K$.
 We also show energies for some configurations for which the two methods
 differ substantially.
 Configurations are distinguished by the shapes of the points they are plotted
 with (those occurring in both panels have the same marker).
 Colours of displayed points refer to the isotonic chain.

\textbf{\textit{ Isotones N=143, 145}}\
 The low-lying high-$K$ $3\nu$ configurations in these isotones
 have to contain at least two of the $\nu5/2^+_{\sf 7}, \nu5/2^-_{\sf 7},
 \nu 7/2^-_{\sf 5}$ or $\nu 7/2^+_{\sf 4}$ orbitals - see Fig. \ref{nlev2}.

 In the q.p. method, the configurations composed of $\nu 5/2^-_{\sf 7}\otimes
 \nu 7/2^-_{\sf 5}$ and either $\nu 1/2^+_{\sf 13}$
  or  $\nu 1/2^-_{\sf 13}$ neutron states are the lowest in lighter $N=143$
 isotones (black squares and crosses in the lower panel of Fig. \ref{3n15143}).
 The configuration with $\nu 5/2^+_{\sf 7}$ instead of
  $\nu 5/2^-_{\sf 7}$ lies $\approx 200$ keV higher (black asterisk in the
  lower panel of Fig. \ref{3n15143}). In the $N=145$ isotones, the lowest
  configuration is
   $\nu^3\{1/2^+_{13}\otimes 5/2^+_{\sf 7}\otimes 7/2^-_{\sf 5}\}$, at
   $\approx 1.3 - 1.5$ MeV in Fm, No, and Rf (red circles in the lower panel of
  Fig. \ref{3n15143}).

  Results of the PNP method are similar, with the same lowest configuration in
  $N=145$ isotones (upper panel of Fig. \ref{3n15143}). The lowest PNP
  $3\nu$ configurations in $N=143$ have the $\nu 5/2^-_{\sf 7}$ neutron
  replaced with $\nu 5/2^+_{\sf 7}$ (black circles and asterisks).


 As the energies of one-neutron states $\nu 7/2^-_{\sf 5}$,
 $\nu 1/2^+_{\sf 13}$, $\nu 1/2^-_{\sf 13}$ are close to each other, within
 $<100$ keV in lighter $N=143, 145$ isotones, energies of $3\nu$
 configurations containing them should be considered relative to those of
 the collective band built on the $\nu 7/2^-_{\sf 5}$ level. Then, the
 excitation energy of, for example, the
 $\nu^3\{7/2^-_{\sf 5}\otimes 1/2^+_{\sf 13}\otimes 5/2^+_{\sf 7}\}$
 configuration should be compared to the rotational energy
 corresponding to 3 (1/2+5/2) $\hbar$ units of the collective
 angular momentum. The latter value is much smaller than the former, so
  a large hindrance of the decay of the lowest $3\nu$ configuration to this
 band is not expected.

 Therefore, we also show energies of
 $\nu^3\{7/2^-_{\sf 5}\otimes 7/2^+_{\sf 4}\otimes 5/2^{\pm}_{\sf 7}\}$
 configurations with larger $K=19/2$ (black triangles and red inverted
 triangels in upper panel, Fig. \ref{3n15143}).
 However, even the lower one, at $\approx$ 2 MeV in $N$=145 isotones,
 still has estimated excitation above the collective rotational band of more
 than 1 MeV.

 In summary, configurations with the same $K$ lie lower in light $N=145$
 isotones and have a better chance of being isomeric there. However, due to
 $\gtrsim 1$ MeV excitation over low-lying rotational $1\nu$ band, there are
 no prominent candidates for $3\nu$ high-$K$ isomers in $N=143, 145$ isotones.


\textbf{\textit{ Isotones N=147, 149}}\
  The configuration of three states below the $N$=152 gap,
 $\nu^3 \{9/2^-_{\sf 3}\otimes 7/2^+_{\sf 4}\otimes 5/2^+_{\sf 7}\}$, is
 a candidate for an isomer in $N=149$ isotones (red squares
 in Fig. \ref{3n15147}).
  As the g.s. $K$ values predicted for $N$=149 are either
  7/2 or 9/2 (cf Fig. \ref{szach1nq}), the extra aligned angular momentum
  projection is either 7 or 6 $\hbar$ units relative to the g.s. rotational
  bands.  From energies $\approx 1.2 - 1.3$ MeV in the q.p.
 method, and more varied $1.1 - 1.7$ MeV in the PNP,
  we estimate the excitation energy of 0.6 - 0.7 MeV above g.s rotational
  band in No and slightly smaller (larger) in Rf in the q.p. (PNP) method.
  The configuration with $K$ larger by one (red diamonds in upper panel of
  Fig. \ref{3n15147}) lies already $\approx$ 0.5 MeV higher which makes it much
  worse candidate for an isomer.

  The lowest configurations for $N=147$ in the q.p. calculations contain
  the $\nu 1/2^+_{\sf 13}$ state (black triangles and inverted triangles in
  the lower panel of Fig.  \ref{3n15147}). As the predicted g.s. $K$-values
  for $N=147$ isotones are either 5/2 or 7/2 (Fig. \ref{szach1nq}), with
  only 3 or 4 units of extra aligned-$K$ and excitation energy of
  $\approx 1.5$ MeV, these are not the best candidates for isomers.
  Therefore, we show PNP results for other, higher-K states
   (upper panel of Fig. \ref{3n15147}): one that is the lowest for $N=149$
  (black squares)i and the $K=19/2$ configuration with $\nu 7/2^-_{\sf 5}$ and
   $\nu 7/2^+_{\sf 4}$ neutrons (black dots). With extra aligned $K$ of
   7 and 8 relative to the $K$=5/2 g.s. in Fm and No, they are not excluded
   as candidates for isomerism, although less favoured than the one in $N=149$.

\textbf{\textit{ Isotones N=151, 153}}\
  In q.p. calculations, the lowest $3\nu$ energies in $N=151$ isotones are
  obtained for configurations built of neutron orbitals below the $N=152$ gap
  (black squares and circles in the lower panel of Fig. \ref{3n15151}). As
   discussed earlier, this results from an improper position of the Fermi
   energy in the q.p. formalism. In the PNP calculations, the same
   configurations lie much higher (black squares, upper panel). The effect
   of the $N=152$ gap in the PNP calculation makes energies of all $3\nu$
   states are rather high.
  For example, the configuration $\nu^3\{11/2^-_{\sf 2}\otimes 9/2^-_{\sf 3}
  \otimes 7/2^+_{\sf 4}\}$ with extra 9 $\hbar$ units of $K$ relative to the
  g.s. (which is predicted as the $\nu9/2^-_{\sf 3}$ state - cf Fig.
  \ref{szach1nq}) has energy above 2 MeV in PNP calculations
  (black triangles in the upper panel). This corresponds to
  $\approx 1.4 (1.)$ MeV estimated excitation above the g.s. rotational band
  in Fm (Sg), which seems large for an isomer.

  Calculated energies of low-lying $3\nu$ configurations in $N$=153 isotones
  are greater than 1.8 (1.7) MeV in the q.p. (PNP) method.
  The similarity of the q.p. results to those of the PNP follows from the
  choice of the $N-1=152$ system as the core: pairing Fermi energy is situated in the middle of the $N=152$ gap, hence energies of quasiparticles from below the gap are not so underestimated as for $N=151$.
  At large excitation energies, interesting configurations with
  large extra aligned $K$ relative to the low-lying $1\nu$ bandheads with
   $K=1/2$ (g.s.), 3/2 and 11/2. The PNP energies for the
$\nu^3\{\nu 11/2^-_{\sf 2}\otimes \nu 9/2^-_{\sf 3}\otimes\nu 1/2^+_{\sf 14}\}$
 and
$\nu^3\{\nu 11/2^-_{\sf 2}\otimes \nu 9/2^-_{\sf 3}\otimes\nu 7/2^+_{\sf 5}\}$
 configurations are shown in Fig. \ref{3n15151}
  (upper panel, red inverted triangles and pentagrams).
  Since the $K=11/2$ band head lies only $\approx 200$ keV above the $K=1/2$
  g.s. in Fm and becomes the g.s. in Ds and Cn, crucial are the excitation
   energies of these $3\nu$ states relative to the rotational
   states with $I=K_{\nu^3}$ built on the $K=11/2$
   and $K=1/2$ band heads.
   The estimated values of those in Fm - Sg are: $\approx 1.1$ MeV for
  the $K=21/2$ configuration (roughly the same relative to both band heads) and
   $\approx 1$ MeV for the $K=27/2$ one
- not excluding, but not favouring their isomerism either.

\textbf{\textit{ Isotones N=155, 157}}\
 Low-lying $3\nu$ states in these isotones are built from neutron
 orbitals above the $N=152$ gap.
  Ground states predicted within the q.p. method are formed by the
 $1/2^+_{\sf 14}$ (in Fm - Hs) and $11/2^-_{\sf 2}$ (in Ds, Cn) orbitals
  (cf Fig. \ref{szach1nq}). These two $1\nu$ configurations and the one built
   on the $\nu 3/2^+_{\sf 10}$ level lie very close to each other in
  $N=155, 157$ isotones (within less than 100 keV, often nearly degenerate).
  In $N=157$ isotones, additionally $\nu 7/2^+_{\sf 5}$ and $9/2^+_{\sf 3}$
  states lie very close to the above (this relates to differences between
  the PNP and q.p. predictions for g.s. configurations in N=157 isotones -
   cf Fig. \ref{szach1nq}).

  In the q.p. method, the lowest $3\nu$ configuration in both isotone chains
  is $\nu^3\{11/2^-_{\sf 2}\otimes 1/2^+_{\sf 14}\otimes 3/2^+_{\sf 10}\}$
  at $E^*_{\nu^3}<$ 1.5 MeV, with lower energies for $N=157$
 (black and red circles in the lower panel of Fig. \ref{3n15155}).
  In the PNP method, this configuration lies very low in $N=155$ isotones.
  Above it, at $E^*_{\nu^3}\approx 1.5$ MeV, there are two states
   $\nu^3\{11/2^-_{\sf 2}\otimes 7/2^+_{\sf 5}\otimes 1/2^+_{\sf 14}
   (3/2^+_{\sf 10})\}$, one of which is shown in Fig. \ref{3n15155} (black
   triangles); at $E^*_{\nu^3}\approx 2$ MeV lie two other $3\nu$ states:
  $\nu^3\{11/2^-_{\sf 2}\otimes 9/2^+_{\sf 3}\otimes 3/2^+_{\sf 10}\}$  and
  $\nu^3\{11/2^-_{\sf 2}\otimes 9/2^+_{\sf 3}\otimes 7/2^+_{\sf 5}\}$
  (black inverted triangles and diamonds). Estimated excitation of these
  states above the rotational band built on the $K=11/2$ band head (at their
  respective angular momentum values) are $\approx 0.9 - 1$, 1.1 and 1.2
  MeV, respectively, for the $K=15/2$, 19/2, and 27/2 configurations.

  In the PNP calculations for $N=157$ isotones, the $\nu^3\{11/2^-_{\sf 2}
  \otimes 1/2^+_{\sf 14} \otimes 3/2^+_{\sf 10}\}$ configuration is not the
  lowest one; its energy (not shown in Fig. \ref{3n15155}) is similar as those
   of two other configurations:
  $\nu^3\{11/2^-_{\sf 2}\otimes 7/2^+_{\sf 5}\otimes 3/2^+_{\sf 10}\}$
  and $\nu^3\{11/2^-_{\sf 2}\otimes 9/2^+_{\sf 3}\otimes 7/2^+_{\sf 5}\}$
  (red inverted trangles and diamonds in the upper panel of Fig. \ref{3n15155}).
  The configurations $\nu^3\{11/2^-_{\sf 2}\otimes 7/2^+_{\sf 5}
 \otimes 1/2^+_{\sf 14} (3/2^+_{\sf 10})\}$ lie lower, at $\approx 1.3 - 1.5$
  MeV (red squares and triangles in the upper panel of Fig. \ref{3n15155}).
  These four configurations lie lower than in $N=155$ isotones.
  Their excitation above the $K$=11/2 g.s. rotational band amounts to
   $\approx 0.8$ MeV for the
   $\nu^3\{11/2^-_{\sf 2}\otimes 7/2^+_{\sf 5}\otimes 3/2^+_{\sf 10})\}$
 state and $\approx 0.9$ MeV for the $K=27/2$ configuration -
  so they are likeky candidates for isomers.


\textbf{\textit{ Isotones N=159, 161, 163}}\
  Energies of low-lying high-$K$ configurations in $N$=159 - 163 even-odd
  isotones are depicted in Fig. \ref{3n15159}, \ref{3n15163}.
  The ground states of these nuclei are predicted by the q.p. calculation
  are mostly formed by the $7/2^+_{\sf 4}$, $9/2^+_{\sf 3}$ and
  (exclusively) $13/2^-_{\sf 1}$ neutron states for $N$=159, 161, 163,
  respectively, while in the PNP method, one often obtains the $K^{\pi}=9/2^+$
  g.s. in $N=159$ isotones (cf Fig \ref{szach1nq}).

  Low-lying $3\nu$ states in $N=159$ isotones are built from neutron
 orbitals below the $N=162$ gap. The best candidate for isomer seems to be
 the $\nu^3\{11/2^-_{\sf 2}\otimes 9/2^+_{\sf 3}\otimes 7/2^+_{\sf 5}\}$
  configuration, at $E^*_{\nu^3} \approx$ 1.3-1.5 MeV in the PNP calculations,
  and 1.2 - 1.3 MeV in the q.p. method (black squares in Fig. \ref{3n15159}),
   with an extra 9 or 10 $\hbar$ units of $K$ with
   respect to the g.s. (which has $K=7/2$ or $9/2$ - Fig. \ref{szach1nq}).
  Its excitation above the same spin state of the g.s.
  rotational band is estimated as 0.3 - 0.4 MeV for $^{263}$Rf.
  For the Ds and Cn, the configuration
  $\nu^3\{3/2^+_{\sf 10}\otimes 9/2^+_{\sf 3}\otimes 7/2^+_{\sf 5}\}$ lies
  lower, but has smaller $K$.

  The same configuration as in $N=159$ is favoured for $N=161$ isotones
  within the q.p.  method. In the PNP calculation, which accounts for
   proper position of the Fermi energy, it is highly  excited
  as all three states lie below the $N$=162 gap (red squares in
  Fig. \ref{3n15159}). The more interesting is
  the optimal configuration $\nu^3\{13/2^-_{\sf 1}\otimes 9/2^+_{\sf 3}
 \otimes 7/2^+_{\sf 5}\}$ (red diamonds in Fig. \ref{3n15159}) that has
  $K$ by 10 units of $\hbar$ bigger than the g.s.
  and energy $E^*_{\nu^3}=1.7 - 2$ MeV, decreasing with $Z$.
  Its estimated excitation above the g.s. rotational band at spin
  $I=29/2$ is $\approx 0.7$ MeV in $^{265}$Rf and less than that
  in Ds and Cn. It seems a very good candidate for the higher-lying high-$K$
  isomer.

  Excitation energies of high-$K$ $3\nu$ configurations in $N$=163 isotones
  are mostly greater than 2 MeV in both variants of calculations
 (due to the choice of $N-1$ system as a core in the q.p. method)
   - Fig. \ref{3n15163}. Still, for example, the energy of the configuration
  $\nu^3 33/2^+\{13/2^-_{\sf 1}\otimes 11/2^-_{\sf 2}\otimes 9/2^+_{\sf 3}\}$
  (triangles, lower panel) is underestimated by the q.p. method by
  0.5 - 1.2 MeV compared to the more reliable PNP calculation.
  Out of two lowest-lying high-$K$ $3\nu$ configurations obtained in the PNP
  method, shown in Fig. \ref{3n15163}, the more interesting is the one
  with higher $K$,
  $\nu^3 31/2^-\{13/2^-_{\sf 1}\otimes 9/2^+_{\sf 4}\otimes 9/2^+_{\sf 3}\}$
 (squares, upper panel). It would lie at an estimated $\approx 1$ MeV above the
  same spin state of the g.s. rotational band built on the predicted $K=13/2$
  ground state. While such excitation is not small, this
  configuration could be a higher-lying, very high-$K$ isomer.


\textbf{\textit{ Isotones N=165, 167}}\
  Ground states in $N=165$ isotones are formed mostly by the
  $13/2^-_{\sf 1}$ orbital, while the $1\nu$ $9/2^+_{\sf 4}$ and
  $3/2^+_{\sf 11}$ configurations lie close to it for, respectively,
   lighter and heavier isotones. For $N=167$ situation is even more
   complicated as closely lying $1\nu$ states: $11/2^+_{\sf 2}$,
  $9/2^+_{\sf 4}$, $3/2^+_{\sf 11}$ and $5/2^+_{\sf 8}$ compete to become
  the g.s., with $13/2^-_{\sf 1}$ state lying close (within 150 keV) up to
  $Z=106$.

 The same high-$K$ $3\nu$ configurations:
 $\nu^3\{13/2^-_{\sf 1}\otimes 5/2^+_{\sf 8} \otimes 3/2^+_{\sf 11}\}$ and
 $\nu^3\{13/2^-_{\sf 1}\otimes 9/2^+_{\sf 4} \otimes 3/2^+_{\sf 11}\}$
 come out as the lowest-lying in both isotone chains in the q.p. calculation
 (black and red circles and triangles, lower panel of Fig. \ref{3n15165}).
 In the PNP calculation, these configurations are also the lowest
 in $N=165$ isotones (with energies 200-300 keV lower than the q.p. results).
 In Fig. \ref{3n15165}  we also show energies for
 $\nu^3\{13/2^-_{\sf 1}\otimes 11/2^+_{\sf 2} \otimes 9/2^+_{\sf 4}\}$ and
 $\nu^3\{13/2^-_{\sf 1}\otimes 9/2^+_{\sf 4} \otimes 5/2^+_{\sf 8}\}$
 states in $N=165$ nuclei (black inverted triangles and squares, upper
  panel of Fig. \ref{3n15165}).

 For $N=167$, PNP energies of configurations favoured in the q.p. calculations
 are much higher, nearly equal to those of the
 $\nu^3\{13/2^-_{\sf 1}\otimes 9/2^+_{\sf 4} \otimes 5/2^+_{\sf 8}\}$ state,
  shown in Fig. \ref{3n15165} (red squares, upper panel).
 Configurations that come out lower in the PNP method are:
 $\nu^3\{9/2^+_{\sf 4}\otimes 5/2^+_{\sf 8} \otimes 3/2^+_{\sf 11}\}$ and
 $\nu^3\{11/2^+_{\sf 2}\otimes 9/2^+_{\sf 4} \otimes 3/2^+_{\sf 11}\}$
 (red diamonds and asterisks, upper panel, Fig. \ref{3n15165}).
 Thus, the q.p. and PNP results suggest a better chance for isomers
 in different isotonic chains.

  For PNP results, estimates of the excitation above the g.s. rotational band
   are as follows. For $N=165$, with the $K=13/2$ g.s., the lowest two
  $3\nu$ states have extra $K$ of 4 or 6 $\hbar$ units.
  Collective rotation reduces excitation $E^*_{\nu^3}$ for Hs by, respectively,
  0.45 and 0.7 MeV, to $\approx 0.8$ MeV in both cases.
  For the configuration with the highest $K=33/2$ (black inverted triangles),
  with extra 10 $\hbar$ units of $K$, the estimate for Hs gives $\approx 0.5$
  MeV above the g.s. band. This makes all these states, and especially one with
  the highest-$K$, good candidates for isomers. The $3\nu$ states in $N=167$
  isotones lie higher than in $N=165$, but still some of them can be
   isomeric.

\textbf{\textit{ Isotones N=169, 171, 173}}\
 In $N=169$ isotones of Hs - Cn the $1\nu$ states formed by the $9/2^+_{\sf 4}$
  and $11/2^+_{\sf 2}$ orbitals are within a few tens of keV from the ground
 state. In Ds and Cn nuclei with $N=171, 173$, their smaller equilibrium
 deformations $\beta_{20}$ make a unique $15/2^-_{\sf 1}$ orbital equally
  close to the neutron Fermi level.
 Therefore one should expect very low-lying high-$K$ $1\nu$ isomers (or ground
 states, as the energy differences of 10 - 30 keV between $1\nu$ configurations
 in the PNP calculation cannot be treated as absolutely certain) in these
 nuclei.

 The lowest-lying high-$K$ $3\nu$ configurations in $N$=169 - 173 isotones are:
  $\nu^3 25/2^+\{5/2^+_{\sf 8}\otimes 11/2^+_{\sf 2}\otimes
  9/2^+_{\sf 4}\}$ at $E^*\approx$ 1.45 MeV
   for $N$=169 in Hs, Ds and Cn,
  $\nu^3 23/2^-\{5/2^+_{\sf 8}\otimes 15/2^-_{\sf 1}\otimes
  3/2^+_{\sf 11}\}$ at $E^*\approx$ 1.3 (1.25) MeV
  for $N$=171 in Ds (Cn) and
  $\nu^3 21/2^-\{5/2^+_{\sf 8}\otimes 15/2^-_{\sf 1}\otimes
  1/2^+_{\sf 15}\}$ at $E^*$=1.36 MeV
  for the $N$=173 isotone of Cn. To appreciate whether there is a hindrance of
  their decay to lower-lying collective structures one should consider
  rotational bands built on the high-$K$ $1\nu$ band heads.

  Considering the abovementioned $3\nu$ configuration in $N=169$ isotones, the
  extra-aligned $K$ with respect to the low-lying band built on the
  $\nu 11/2^+_{\sf 2}$ state of 7 $\hbar$ units points to $\approx 0.6$ MeV
  of excitation over the collective structure and a good chance for isomerism.
  Regarding $3\nu$ configurations in $N=171, 173$ nuclei one encounters two
  counteracting circumstances: 1) their extra-aligned $K$ with respect to the
  low-lying $1\nu$ band built on the $\nu 15/2^-_{\sf 1}$ state,
 of 4 and 3 $\hbar$, respectively, is not large, 2) a reduced deformation
  $\beta_{20}$ increases rotational energies.
  As the energies of these $3\nu$ configurations are rather low, we expect that
  they also have a quite good chance of being isomeric.


\subsection{Comparison with experimental evidence}

 Regarding a comparison of the above results with the experimental evidence,
 one can start with the very low-lying $11/2^-$ isomers
 found in $^{253}$Fm (at $\approx$ 350 keV) \cite{Antalic2011}, $^{255}$No
 (at 240-300 keV \cite{Bronis2022} or 225 keV \cite{Kessaci2024})
 and $^{257}$Rf (at $\approx$ 75 keV) \cite{Berr2010}, attributed to the
 $\nu 11/2^-_{\sf 2}[725]$ s.p. state. In the PNP calculation we obtain
 for those nuclei the following excitation energies of this configuration:
 280, 240 and 180 keV, in a reasonable agreement with the data.

 Two 3qp isomers were reported in $N=$149 isotones: in $^{251}$No
 ($\approx 2\mu$s, at $> 1.7$ MeV) \cite{Hess2006,Lop2022}
 and in $^{253}$Rf (0.66 ms at $\gtrsim$
 1.02 MeV \cite{Lop2022} or $\approx$ 0.6 $\mu$s \cite{Khuy2021}).
 In the q.p. method, the $1\nu2\pi$ excitations in $^{251}$No including the
 lowest neutron $7/2^+_{\sf 4}$ state have the following energies and
 proton contents: 1.467 MeV $\pi^2\{9/2^+_{\sf 2}\otimes 1/2^-_{\sf 10}\}$,
  1.554 MeV $\pi^2\{9/2^+_{\sf 2}\otimes 7/2^-_{\sf 3}\}$, and
  1.807 MeV $\pi^2\{7/2^+_{\sf 3}\otimes 7/2^-_{\sf 3}\}$. The 3 q.p.
 configurations with the neutron state changed to $\nu 9/2^-_{\sf 3}$ and
 $\nu 5/2^+_{\sf 7}$ lie $\approx$ 10 and 70 keV above them, respectively.
 In the PNP calculations, the energy 1.442 MeV of the lowest configuration is
 similar to the above, but that of the one with the $\pi^2 8^-$ pair is
  higher, $E^*=1.85$ MeV.  The lowest $3\nu$ q.p. state in $^{251}$No,
   $\nu^3\{9/2^-_{\sf 3}\otimes 7/2^+_{\sf 4}\otimes 5/2^+_{\sf 7}\}$,
  has even lower energy: 1.172 MeV in the q.p. method and 1.149 MeV in PNP
 (red squares in Fig. \ref{3n15147}).
%
%
  The two lowest $1\nu2\pi$ configurations in $^{253}$Rf obtained within the
  q.p. method contains the proton  $\pi^2 8^-\{9/2^+_{\sf 2}\otimes 7/2^-_{\sf 3}\}$ pair and either $\nu 9/2^-_{\sf 3}$ at 1.222 MeV or $\nu 7/2^+_{\sf 4}$ at 1.247 MeV;
  the PNP energies of those states are: 1.013 and 1.093 MeV, respectively.
  The same as in $^{251}$No lowest $3\nu$ configuration is calculated at
  1.222 MeV in the q.p. method and at 1.295 MeV in PNP.
  From the above results, one would expect in $^{251}$No an isomer at energy
  significantly lower than 1.7 MeV, whereas calculated energies of $1\nu2\pi$
  and $3\nu$ isomers in $^{253}$Rf roughly agree with the one suggested in
 \cite{Lop2022}.

  In $N=151$ isotones, one 3q.p. isomer ($I\ge 23/2$, 0.63 ms, $E^*>1.44$ MeV)
 is known in $^{253}$No \cite{Lop2007,Streicher2010,Antalic2011} and two
 ($I^{\pi}=19/2^+$, 29 $\mu$s, $ E^*$=1.103 MeV and $I^{\pi}=25/2^+$,
 49 $\mu$s, $E^*$=1.303 MeV) in $^{255}$Rf \cite{Mosat2020,Chakma2023}.
 The isomer in $^{253}$No is attributed either to one of the
   $\nu^3\{9/2^-_{\sf 3}\otimes 7/2^+_{\sf 4}\otimes 7/2^+_{\sf 5}\}$ or
   $\nu^3\{9/2^-_{\sf 3}\otimes 5/2^+_{\sf 7}\otimes 7/2^+_{\sf 5}\}$
 structures \cite{Lop2007}, or to the neutron $9/2^-[734]$ state coupled
 to two-proton $\pi^2 \{9/2^+[624]\otimes 7/2^-[514]\}$ configuration
 \cite{Streicher2010,Antalic2011}.
 Two isomers in $^{255}$Rf are attributed to the neutron state $9/2^-[734]$
 coupled with the $\pi^2\{9/2^+[624]\otimes 1/2^-[521]\}$ and
 $\pi^2\{9/2^+[624]\otimes 7/2^-[514]\}$ proton pair
  \cite{Mosat2020,Chakma2023}. Results of our calculations point to their
  $1\nu2\pi$ rather than $3\nu$ structure, as follows from the PNP results,
    more reliable for $N$=151.
  In particular, the calculated energy for the candidate $3\nu$ state
   $\nu^3\{9/2^-_{\sf 3}\otimes 7/2^+_{\sf 4}\otimes 7/2^+_{\sf 5}\}$
 in $^{253}$No is above 2 MeV (cf also neutron levels for $^{253}$No in
  Fig. \ref{nlev2}).
 Thus, the likely structure of these isomers seems to be as suggested in
 \cite{Chakma2023} for $^{255}$Rf.
  Calculated energies of both $1\nu2\pi$ configurations in $^{253}$No
 are somewhat high: 1.44 and 1.89 MeV (in the PNP method), and
 1.44 and 1.53 MeV (in the quasiparticle method).
 In $^{255}$Rf, their energies: 1.026 and 1.434 MeV (in PNP), or 1.20 and 1.425
  MeV (in the q.p. method) well agree with the data. However,
 the main problem in $^{255}$Rf is the unlikely higher position of the state
containing the $K_{2\pi}=5$ proton pair relative to that with the $K_{2\pi}=8$.

  In $N=153$ isotones, two 3q.p. isomers are known in $^{255}$No
 ($I=$19/2 - 23/2 at 1.4 - 1.6 MeV and $I\ge 19/2$ at $\ge$1.5 MeV according
 to \cite{Bronis2022} or $I^{\pi}=21/2^+$ at $\approx 1.36$ MeV and
 $27/2^+$ at $\approx$ 1.5 MeV according to \cite{Kessaci2024})
 and one (106 $\mu$s, $I^{\pi}=21/2^+$ at 1.151 MeV
 \cite{Qian2009,Berr2010,Riss2013}) in $^{257}$Rf.
  The structure suggested for two $^{255}$No isomers in \cite{Kessaci2024}
 is: $\nu 11/2^-[725]$ state coupled to $\pi^2\{1/2^-[521]\otimes 9/2^+[624]\}$
  and $\pi^2\{7/2^-[514]\otimes 9/2^+[624]\}$. The isomer in $^{257}$Rf
  was attributed to the first of the above configurations
  \cite{Berr2010,Riss2013}.
  Based on our results, $3\nu$ states in $N=153$ isotones have energies greater
 than 1.7 MeV (Fig. \ref{3n15147}).  Among predicted $1\nu2\pi$ configurations,
  the candidates suggested in \cite{Kessaci2024} for $^{255}$No come out at
  1.74 and 2.12 MeV in PNP
  calculations (1.67 and 1.74 MeV in the quasiparticle method), which is too
  high. In $^{257}$Rf, as in $^{255}$Rf, the suggested configuration is
  predicted unnaturally above the one with the $K_{\pi}=8$ pair: the energy of the latter is 1.24 MeV in PNP and 1.36 in the q.p. method, while that of the former is 1.63 MeV in PNP and 1.59 MeV in the q.p. method.

 A better agreement with experiment for $N$=151 and 153
 isotones of No and Rf could be reached by taking into account
  the inverted order of proton $1/2^-_{\sf 10}$ and $7/2^-_{\sf 3}$ levels.
 As argued in the discussion of $1\nu2\pi$ results, such
 a change, including a smaller gap beteen these two levels than in the present
 Woods-Saxon spectrum would lead to a smaller energy difference between
  configurations including $K_{\pi}$=5 and 8 proton pairs and lower
  energies of $1\nu2\pi$ states in No isotopes.

\section{CONCLUSIONS}

 This work presents a systematic theoretical search for possible
 three-quasiparticle ($1\nu2\pi$ and $3\nu$) high-$K$ isomers in even-odd
 Fm--Cn nuclei within a Microscopic-Macroscopic model with the deformed
 Woods-Saxon potential \cite{Jachimowicz2021}. 

 We performed two sets of calculations differing in the treatment of pair correlations: 
one using the quasiparticle method and the other using the particle number projection. 
Both give similar results for excitation energies of $1\nu-2\pi$ high-$K$ states 
and point to mostly the same configurations as candidates for isomers. 
Deficiencies of the simpler quasiparticle approach show up in calculated excitation energies 
of $3\nu$ states, which are particularly unreliable for non-optimal configurations 
at neutron numbers close to the large gaps in the single-particle spectrum, 
i.e., for $N=N_{\text{gap}}\pm1$ (where $N_{\text{gap}}$ denotes the gap position).

Candidates for three-quasiparticle (3q.p.)\ high-$K$ isomers were selected on the basis 
of their low excitation energy and the small estimated energy above the collective 
rotational sequence built on the one-quasiparticle (1q.p.)\ band head. The latter 
involves one of its components (the $1\nu$ state in the case of $1\nu-2\pi$ 
configurations, or any of the three $1\nu$ states in the case of the $3\nu$ 
configuration). The second estimate, based on cranking calculations~\cite{momJ}, 
is rather schematic, a limitation dictated by the scarcity of available 
experimental data.

The most promising candidates for high-$K$ $1\nu-2\pi$ isomers, with excitation 
energies as low as $\approx 1.0$~MeV, are predicted in Rf~($Z=104$) and 
Sg~($Z=106$) isotopes for neutron numbers up to $N \approx 159$. 
Relatively low-energy candidates occur in Ds around $N=163$ and, with 
slightly higher energy and higher $K$, for $N=157$--$161$.

 (involving the $K_{2\pi}=10$ proton pair).
 In Hs ($Z=108$), where the proton gap increases the lowest excitation energies,
 and in experimentally accessible Cn nuclei, with configurations having
 moderate aligned $K_{2\pi}=6$, isomerism is possible. In No nuclei,
 best candidates occur for $N=149 - 155$, while there are not many
  candidates in Fm.

 Best candidates for the $3\nu$ high-$K$ isomers following from the more
 reliable PNP results occur in isotones $N=149, 157, 159$ and 165.
  The energy penalty for promoting a neutron pair across the $N=152$ or
 $N=162$ gap rises energies of $3\nu$ configurations in $N=151, 153, 161$ and
 163 isotones. However, for $N=161$ and 163, our model predicts higher-lying configurations with $K\approx 15$, which could have some hindrance to EM decay.
 Furthermore, for the heaviest isotopes ($N \ge 169$),
  relatively small excitation energies and reduced quadrupole deformation
 which increases rotational energies, could create conditions for isomeric
  state.

 Comparisons of our results with available experimental data on ground states,
 1 q.p. and 3q.p. isomers (in No and Rf isotopes) show some agreement
 but also some differences. In particular, seemingly contrary to evidence,
 the calculation predicts the $1\nu2\pi$ state with the $K_{2\pi}=5$ proton
 pair above the one with the $K_{2\pi}=8$ one in $^{255,257}$Rf, the reported
 g.s. in $^{259}$No has spin/parity $9/2^+$ vs the predicted $11/2^-$, and the
 calculated position of the $1\nu2\pi$ configurations with the neutron
 $11/2^-_{\sf 2}[725]$ orbital in $^{255}$No and $^{257}$Rf is higher
 than suggested by their isomeric attribution. As discussed in the text,
 a part of these discrepancies may result from the relative position of proton
 $1/2^-_{10}$ and $7/2^-_{\sf 3}$ orbitals in our Woods-Saxon potential.

 In conclusion, in view of the structure of the deformed mean field in superheavy
 nuclei it is nearly certain that high-$K$ isomers will be found in many
 even-odd isotopes of the studied region.
 Predictions of our MM model will be confronted with forthcoming experimental
 results and in this way, many features will be tested. In particular,
 the presence of the $N=152$ and $N=162$ gaps in the s.p. neutron spectrum
 would be incompatible with the
 existence of low-lying $3\nu$ isomers in $N=151$, 152, 161, and 163 isotones.
 This and other questions await resolution, and this emphasizes the need for new
  data to test and constrain the fundamental aspects of nuclear models.

\section*{ACKNOWLEDGEMENTS}

M.~K. was co-financed by the International Research Project COPIGAL.


\end{document}